\documentclass[usenatbib]{mn2e}

\usepackage{epsf}
\usepackage{graphicx}
\def\gsim { \lower .75ex \hbox{$\sim$} \llap{\raise .27ex \hbox{$>$}} }
\def\lsim { \lower .75ex \hbox{$\sim$} \llap{\raise .27ex \hbox{$<$}} }
\newcommand{\rms}{\sigma_{\rm fit}}

\begin{document}

\title[CDM Halo Structure]{The statistics of $\Lambda$CDM Halo
Concentrations} \author[Neto et al.] {
\parbox{\textwidth}{Angelo F. Neto$^{1,2}$\thanks{E-mail: fausti@if.ufrgs.br}, Liang Gao$^2$, Philip
Bett$^2$, Shaun Cole$^2$, Julio F. Navarro$^{2,3}$\thanks{Fellow of
the Canadian Institute for Advanced Research}, Carlos S. Frenk$^2$,
Simon D.M. White$^4$, Volker Springel$^4$, Adrian Jenkins$^2$}
\vspace*{4pt} \\
$^1$Instituto de F\'{i}sica, Universidade Federal do Rio Grande do Sul, Porto Alegre RS, Brazil \\
$^2$Institute of Computational Cosmology, Department of Physics, University of Durham,\\ Science Laboratories, South Road, Durham DH1 3LE, UK \\
$^3$Department of Physics and Astronomy, University of Victoria, PO 
Box 3055 STN CSC, Victoria, BC, V8W 3P6 Canada \\
$^4$Max-Planck Institute for Astrophysics, Karl-Schwarzschild Str. 1,
D-85748, Garching, Germany
}
\maketitle

\begin{abstract}
We use the {\it Millennium Simulation} ({\sl MS}) to study the
statistics of $\Lambda$CDM halo concentrations at $z=0$. Our results
confirm that the average halo concentration declines monotonically
with mass; a power-law fits well the concentration-mass relation for
over 3 decades in mass, up to the most massive objects to form in a
$\Lambda$CDM universe ($\sim 10^{15} h^{-1} \, M_{\odot}$). This is in
clear disagreement with the predictions of the model proposed by
\citeauthor{bullock01} for these rare objects, and agrees better with the
original predictions of \citeauthor*{nfw97}. The large volume
surveyed, together with the unprecedented numerical resolution of the
{\sl MS}, allow us to estimate with confidence the distribution of
concentrations and, consequently, the abundance of systems with
unusual properties.  About one in a hundred cluster haloes ($M_{200} \,
\gsim \, 3\times 10^{14}\, h^{-1} \, M_{\odot}$) have concentrations
exceeding $c_{200}=7.5$, a result that may be used to interpret the
likelihood of unusually strong massive gravitational lenses, such as
Abell 1689, in the $\Lambda$CDM cosmogony. A similar fraction (1 in
100) of galaxy-sized haloes ($M_{200}\sim 10^{12} \, h^{-1} \,
M_{\odot}$) have $c_{200}<4.5$, an important constraint on models that
attempt to reconcile the rotation curves of low surface-brightness
galaxies by appealing to haloes of unexpectedly low concentration.  We
find that halo concentrations are independent of spin once haloes
manifestly out of equilibrium are removed from the sample.  Compared
to their relaxed brethren, the concentrations of out-of-equilibrium
haloes tend to be lower and to have more scatter, while their spins
tend to be higher. A number of previously noted trends within the halo
population are induced primarily by these properties of unrelaxed
systems. Finally, we compare the result of predicting halo
concentrations using the mass assembly history of the {\it main
progenitor} with estimates based on simple arguments based on the
assembly time of {\it all progenitors}. The latter typically do as
well or better than the former, suggesting that halo
concentration depends not only on the evolutionary path of a halo's
main progenitor, but on how and when {\it all} of its constituents
collapsed to form non-linear objects.
\end{abstract}
\begin{keywords}
cosmology: theory, cosmology: dark matter, galaxies: haloes, methods: numerical
\vspace*{-0.5 truecm}
\end{keywords}

% The width of this distribution is clearly
%related to variations in formation time: haloes that collapse earlier
%than average have higher-than-average
%concentrations. 

%At given mass, the concentration of
%equilibrium haloes follow an approximately lognormal distribution with
%$\sigma_{\log{c}}\sim 0.1$.

%In particular, we examine the dependence of the halo characteristic
%density and concentration on mass, formation time, spin, and
%departures from equilibrium. 

%and that there is a hint that concentrations converge to a constant
%value for very massive systems.

%
%Out-of-equilibrium haloes have a biased and broader
%distribution of concentrations, as well as larger spins, implying that
%care must be exercised in order not to obscure or mistake subtle
%trends between halo properties with those driven by the presence of
%unrelaxed systems.

%\newpage
%\newpage

\section{Introduction}
\label{sec:intro}

The advent of large cosmological N-body simulations has enabled
important progress in our understanding of the structure of dark
matter haloes. As a result, over the past few years a broad consensus has
emerged about the mass function of these collapsed structures, about
their mass profiles and shapes, and about the presence of substructure
within the vitalised region of a halo. The spherically-averaged halo
mass profiles are of particular interest, not only because of their
immediate applicability to a host of observational diagnostics, such
as gravitational lensing and disk galaxy rotation curves, but also
because of the apparent simplicity of their structure.

\citet[hereafter NFW]{nfw95,nfw96,nfw97} argued that
the density profile of a dark matter halo may be approximated by a
simple formula with two free parameters;
\begin{equation}
\frac{\rho(r)}{\rho_{\rm crit}}= \frac{\delta_{\rm c}
}{(r/r_{\rm s})(1+r/r_{\rm s})^2},
\label{eq:nfw}
\end{equation}
where $\rho_{\rm crit}=3H_0^2/8\pi G$ is the critical density for
closure{\footnote{We express the present-day value of Hubble's
constant as $H_0=100 \, h$ km s$^{-1}$ Mpc$^{-1}$.}}, $\delta_c$ is a
characteristic density contrast, and $r_s$ is a scale
radius. Remarkably, this formula seems to hold for essentially all
haloes assembled hierarchically and close to virial equilibrium,
regardless of mass and of the details of the cosmological model. The
cosmological information is encoded in correlations between the
parameters of the NFW profile, so that observational constraints on
such parameters may be translated directly into interesting
constraints on cosmological parameters.

As discussed by NFW, such correlations arise because the
characteristic density of a system appears to evoke the density of the
universe at a suitably defined time of collapse. This result has been
revisited and confirmed by a rich literature on the topic \citep[see, e.g.][]{kravtsov97,avila99,jing00,ghigna00,klypin01,bullock01,eke01}, which has led to the development of a number of
semi-analytic and empirical procedures to explain and predict the
structural parameters of cold dark matter (CDM) haloes as a function of
mass, redshift, and cosmological parameters. The various approaches
differ in detail and lead to significantly different predictions,
especially when extrapolated to halo masses or to redshifts which were
not well sampled by the numerical data on which they were based.

NFW, for example, proposed that the characteristic density is set at
the time when most of the mass of a halo is in non-linear, collapsed
structures. \citet[B01]{bullock01}, on the other hand, argued that
a better fit to their N-body results is obtained by assuming that the
scale radius of haloes of fixed virial mass is independent of redshift,
leading to a substantially different redshift evolution of the halo
structural parameters than envisioned by NFW.  This conclusion was
seconded by \citet[ENS]{eke01} who proposed a
modification of B01's approach to take into account models with
truncated power spectra, such as expected, for example, in a warm dark
matter-dominated universe.

The predictions of these models also differ significantly for
extremely massive haloes, but these predictions have been notoriously
difficult to validate since these systems are woefully
under-represented in simulations that survey a small fraction of the
Hubble volume. For example, NFW argued that the characteristic density
of a halo is set by the mean density of the Universe at the time of
collapse. Very massive systems have, by necessity, been assembled
quite recently (indeed, they are assembling {\it today}), and
therefore they should all have similar characteristic densities. B01's
model, on the other hand, defines collapse redshifts in a different
way, predicting much lower concentrations for very massive objects.

Despite the (qualified) success of these models at reproducing the
{\it average} mass and redshift dependence of halo structural
parameters, they are in general unable to account for the sizable
scatter about the mean relations. As first discussed by \citet{jing00}, a
sizable spread in concentration (of order $\sigma_{\log_{10}{c}}\sim 0.1$)
is seen at all halo masses, and a number of models have attempted to
reproduce this result using semi-analytic models. Amongst the most
successful are those that ascribe variations in concentration to
disparities in the assembly history of haloes of given mass. For
example, \citet[W02]{wechsler02} identify the scatter with
variations in the time when the rate of mass accretion onto the main
progenitor peaks. A similar proposal was advanced by \citet[Z01]{zhao03a,zhao03b}, who argued that the concentration of a halo is effectively
set during periods when the most massive progenitor is in a phase of
fast mass accretion.

Given the disparities between models, it is important to validate
their predictions in a regime different from that used to calibrate
their parameters. Indeed, with few exceptions, most of these studies
have explored numerically a relatively narrow range of halo mass and
redshift, favouring (because they are easier to simulate) haloes with
masses of the order of the characteristic non-linear mass, $M_*$, and
redshifts close to the present day ($z\sim 0$).  Testing these
predictions on a representative sample of haloes of mass much greater
or much lower than $M_*$, or at very high redshift, requires either
simulations of enormous dynamic range, or especially designed sets of
simulations that probe various mass or redshift intervals one at a
time.

One version of the latter approach was adopted by NFW, who simulated
{\it individually} haloes spanning a large range in mass. The price
paid is the relatively few haloes that can be studied using such
simulation series, as well as the lingering possibility that the
procedure used to select the few simulated haloes may introduce some
subtle bias or artifact. Recently, \citet[M07]{maccio07}, have combined
several simulations of varying mass resolution and box size in order
to try and extend earlier results to $M\ll M_*$ scales. This approach
is not without pitfalls, however. For example, in order to resolve a
statistically significant sample of haloes with masses as low as
$10^{10} \, h^{-1} \, M_\odot$, M07 use a simulation box as
small as $14.2 \, h^{-1}$ Mpc on a side, leading to concerns that the
substantial large-scale power missing from such small periodic
realisation may unduly influence the results.

One way to overcome such shortcomings is to increase the dynamic range
of the simulation, so as to encompass a volume large enough to be
representative while at the same time having enough mass resolution to
extend the analysis well below or well above $M_*$. This is the
approach we adopt in this paper, where we use the {\it Millennium
Simulation} ({\sl MS}) to address these issues. The enormous volume of
the {\sl MS} ($500^3 h^{-3}$ Mpc$^{3}$), combined with the vast number
of particles ($2160^3$), make this simulation ideal to characterise,
with minimal statistical uncertainty, the dependence of the structural
parameters of $\Lambda$CDM haloes on mass, spin, formation time, and
departures from equilibrium. We extend and check our {\sl MS} results at low
masses by using an additional large simulation of a $100 \, h^{-1}{\rm
Mpc}$ region with about 10 times better mass resolution.

For reasons discussed in detail below, numerical limitations impose a
lower mass limit of about $10^{12} h^{-1} \, M_\odot$ in our
analysis. Thus, our study does not extend to halo masses as low as
those probed by M07 (whose smallest box is filled with
particles of mass $1.4 \times 10^7 \, h^{-1}\, M_\odot$), but is aimed
at extending previous work to give reliable and statistically robust
results for large and representative samples of haloes over the full
range from $10^{12}$ to $10^{15}\, h^{-1} M_\odot$. The large number
of haloes in the {\sl MS} also allows us to study in detail deviations
from the mean trends and, in particular, the possible presence of
systems with unusual properties, such as clusters with unusually high
concentrations or galaxy haloes of unusually low density. In this paper
we concentrate on the properties of haloes at $z=0$. A second paper
will extend these results to high redshift.

The plan for this paper is as follows. We describe briefly the
simulation and the halo identification technique in
Section~\ref{sec:haloes}. After a brief introduction to the {\sl MS}
(Section~\ref{ssec:ms}) we describe our halo identification
(Section~\ref{ssec:hid}) and selection (Section~\ref{ssec:hsel}) techniques. We also describe the merger trees in Section~\ref{ssec:mergertrees} and the NFW profile fitting procedure in Section~\ref{ssec:proffit}. The
dependence of the halo structural parameters on mass, spin, and
formation time, as well as the performance of various semi-analytic
models designed to predict halo concentrations, are discussed in
Section~\ref{sec:res}. We conclude with a brief summary in
Section~\ref{sec:conc}.

\section{Haloes in the Millennium Simulation}
\label{sec:haloes}

Our analysis is mainly based on haloes identified in the {\it Millennium
Simulation} ({\sl MS}), \citep{springel05a}, and in this section we
describe briefly our halo identification and cataloguing
procedure. For completeness, we begin with a brief summary of the main
characteristics of the {\sl MS}, and then move on to a fairly detailed
characterisation of the halo sample. Readers less interested in these
technical details may wish to gloss over this section and skip to
Section~\ref{sec:res}, where our main results are presented and discussed.

\subsection{The simulations}
\label{ssec:ms}

The {\it Millennium Simulation} is a large N-body simulation of the
concordance $\Lambda$CDM cosmogony. It follows $N=2160^3$ particles in
a periodic box of $L_{\rm box}=500 \, h^{-1}{\rm Mpc}$ on a side. The
cosmological parameters were chosen to be consistent with a combined
analysis of the 2dFGRS \citep{colless01, percival01} and first year
WMAP data \citep{spergel03}. They are $\Omega_{\rm m}=\Omega_{\rm
dm}+\Omega_{\rm b}=0.25$, $\Omega_{\rm b}=0.045$, $h=0.73$,
$\Omega_{\rm \Lambda} = 0.75$, $n=1$, and $\sigma_8=0.9$. Here
$\Omega$ denotes the present day contribution of each component to the
matter-energy density of the Universe, expressed in units of the
critical density for closure, $\rho_{\rm crit}$; $n$ is the spectral
index of the primordial density fluctuations, and $\sigma_8$ is the
linear rms mass fluctuations in $8 h^{-1}$ Mpc spheres at
$z=0$. Compared with the parameter values now favoured by the
three-year WMAP analysis \citep{spergel06}, the main differences are
that a modest tilt, $n=0.95 \pm 0.02$ and a lower $\sigma_8=0.74\pm
0.05$ are favoured by the analysis of these latest data.

With our choice of cosmological parameters, the particle mass in the
{\sl MS} is $8.6 \times 10^8 h^{-1}{\rm M_{\odot}}$. Particle pairwise
interactions are softened on scales smaller than (Plummer-equivalent)
$\epsilon=5\, h^{-1}{\rm kpc}$. Since galaxy-sized haloes ($M\sim
10^{12} \, h^{-1} \, M_\odot$) in the {\sl MS} are represented with
only about $1000$ particles, we have verified that our results are
insensitive to numerical resolution by comparing them with a second
simulation of a smaller volume, $L_{\rm box}=100 \, h^{-1}{\rm Mpc}$, but
of $9$ times higher mass resolution. This simulation adopted the same
cosmological model as the {\sl MS}, and evolved $N=900^3$ particles of
mass $9.5 \times 10^7 h^{-1}{\rm M_{\odot}}$, softened on scales
smaller than $\epsilon=2.4\, h^{-1}{\rm kpc}$.

Both simulations were performed with a special version of the GADGET-2
code \citep{springel05b} that was specially designed for massively
parallel computation and for low memory consumption, a prerequisite
for a simulation of the size and computational cost of the {\sl MS}.

\subsection{Halo identification}
\label{ssec:hid}

The simulation code produced on the fly a {\it friends of friends}
\citep{davis85} (FOF) group catalogue with link parameter, $b=0.2$,
and at least 20 particles per group. At $z=0$, this procedure
identifies $1.77 \times 10^6$ groups in the {\sl MS}. We have also
used SUBFIND, the {\it subhalo finder} algorithm described in
\cite{springel01}, in order to clean up the group catalogue of
loosely-bound FOF structures, and to analyse the substructure within
each halo (see Sec.~\ref{ssec:hsel}).

Like FOF, SUBFIND keeps only substructures containing $20$ or more
particles. In this way, each FOF halo is decomposed into a {\it
background halo} (or the most massive ``substructure'') and zero or
more embedded substructures.  In the {\sl MS}, SUBFIND finds at $z=0$
a total of $1.82\times 10^7$ substructures, with the largest FOF group
containing $2328$ of them.

\subsubsection{Halo centring}
\label{sssec:centre}

Since much of our analysis deals with radial profiles, it is important
to define carefully the centre of a halo. We choose the
position, ${\mathbf r}_{\rm c}$, of the particle with minimum
gravitational potential in the most massive substructure (the potential
centre). Although this seems like a sensible choice, it is important
to check that other plausible options do not lead to large differences
in the location of the halo centre. 

We have therefore compared the potential centre with the result of
the ``shrinking sphere'' algorithm \citep{power03}, which is intended
to converge towards the density maximum of the most massive
substructure, independent of the SUBFIND algorithm. It starts by
enclosing all FOF particles within a sphere and computes iteratively
their centre of mass, shrinking the radius of the sphere by
$r_i=r_0(1-0.025)^i$, and rejecting particles outside the sphere. The
iteration stops when the shrinking sphere contains $1\%$ of the
initial number of particles.

We carried out this comparison in a sub-volume of the {\sl MS}
containing $~2000$ haloes with $N_{\rm FOF}>450$. For $93\%$ of these haloes
the methods agree in the centre position, with a difference smaller
than the gravitational softening, $\epsilon$. However,
we note that the result can depend on the geometry of the FOF group.
When the FOF halo is double (or multiple) and its centre of mass is
far from the centre of the most massive substructure, the shrinking
sphere may converge to another slightly less massive substructure. We
conclude that the potential centre is a more robust determination of
the halo centre, but that the discrepancy between the two methods could be
used to flag problematic haloes whose mass distributions deviate
significantly from spherical symmetry.

\subsubsection{Halo boundary}
\label{sssec:rvir}

Using the potential centre, we define the limiting radius $r_{\rm
lim}$ of a halo by the radius that contains a specified density
contrast $\overline{\rho}(r)=\Delta \, \rho_{\rm crit}$. This defines
implicitly an associated mass for the halo through
\begin{equation}
M=\frac{4}{3}\pi \Delta \, \rho_{\rm crit}r_{\rm lim}^3.
\label{eq:rvir}
\end{equation}
We note that this includes all the particles inside this spherical
volume, and not only the particles grouped by the FOF or the SUBFIND
algorithms.

The choice of $\Delta$ varies in the literature, with some authors
using a fixed value, such as NFW, who adopted $\Delta=200$, and
others, such as B01, who choose a value motivated by the spherical
collapse model, where $\Delta\sim 178\, \Omega_{\rm m}^{0.45}$ (for a
flat universe), which gives $\Delta=95.4$ at $z=0$ for our adopted
$\Lambda$CDM parameters \citep[e.g.][]{lahav91,eke96}. The drawback of
the latter choice is its dependence on redshift and cosmological
parameters. We keep track of {\it both} definitions in our halo
catalogue, but will quote mainly values adopting $\Delta={200}$. When
necessary, we shall specify the choice by a subscript; e.g., $M_{200}$
and $r_{200}$ are the mass and radius of a halo adopting
$\Delta={200}$; $M_{\rm vir}$ and $r_{\rm vir}$ correspond to adopting
$\Delta=95.4$. Unless otherwise specified, quantities listed without
subscript throughout the paper assume $\Delta={200}$.

%%%%%%%%%%%%%%%%%%%%%%%%%%%%%%%%%%%%%%%%%%%%%%
\begin{figure}
\begin{center}
\mbox{ 
\hspace{0.6cm}
\resizebox{3.65cm}{!}{\includegraphics{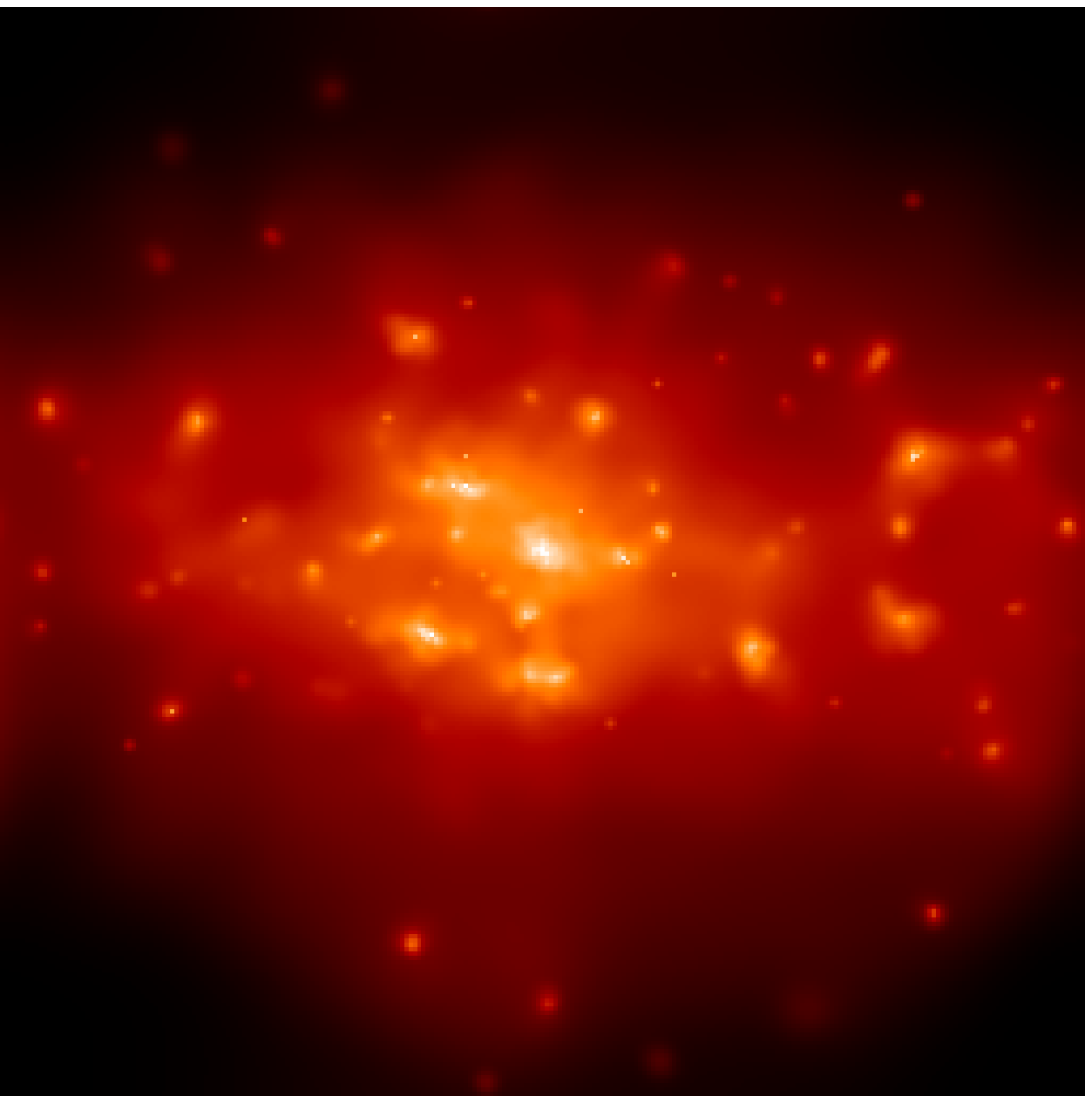}}
\hspace{-0.375cm}
\resizebox{3.65cm}{!}{\includegraphics{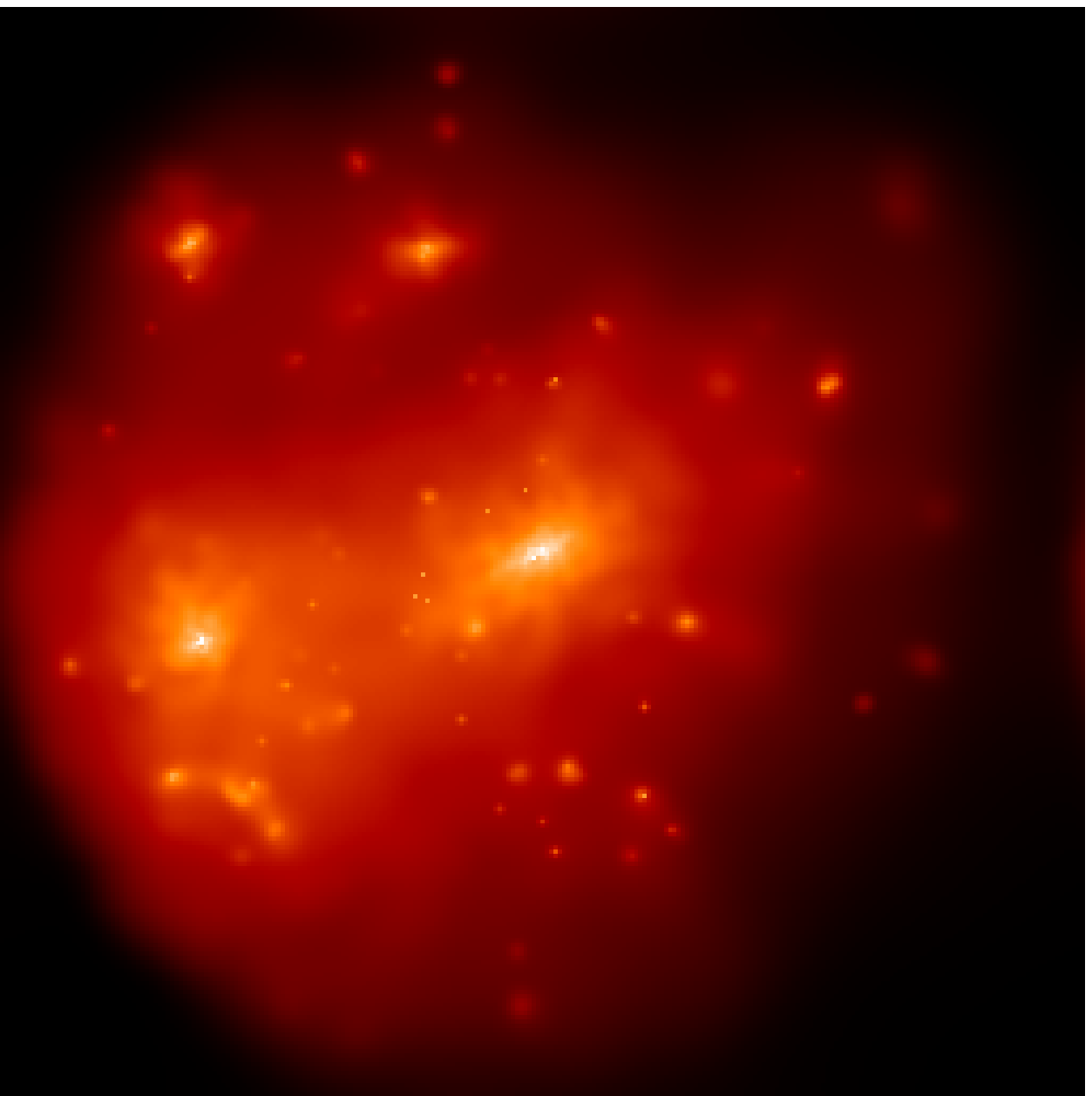}}
}
\mbox{
\hspace{0.6cm}
\resizebox{3.64cm}{!}{\includegraphics{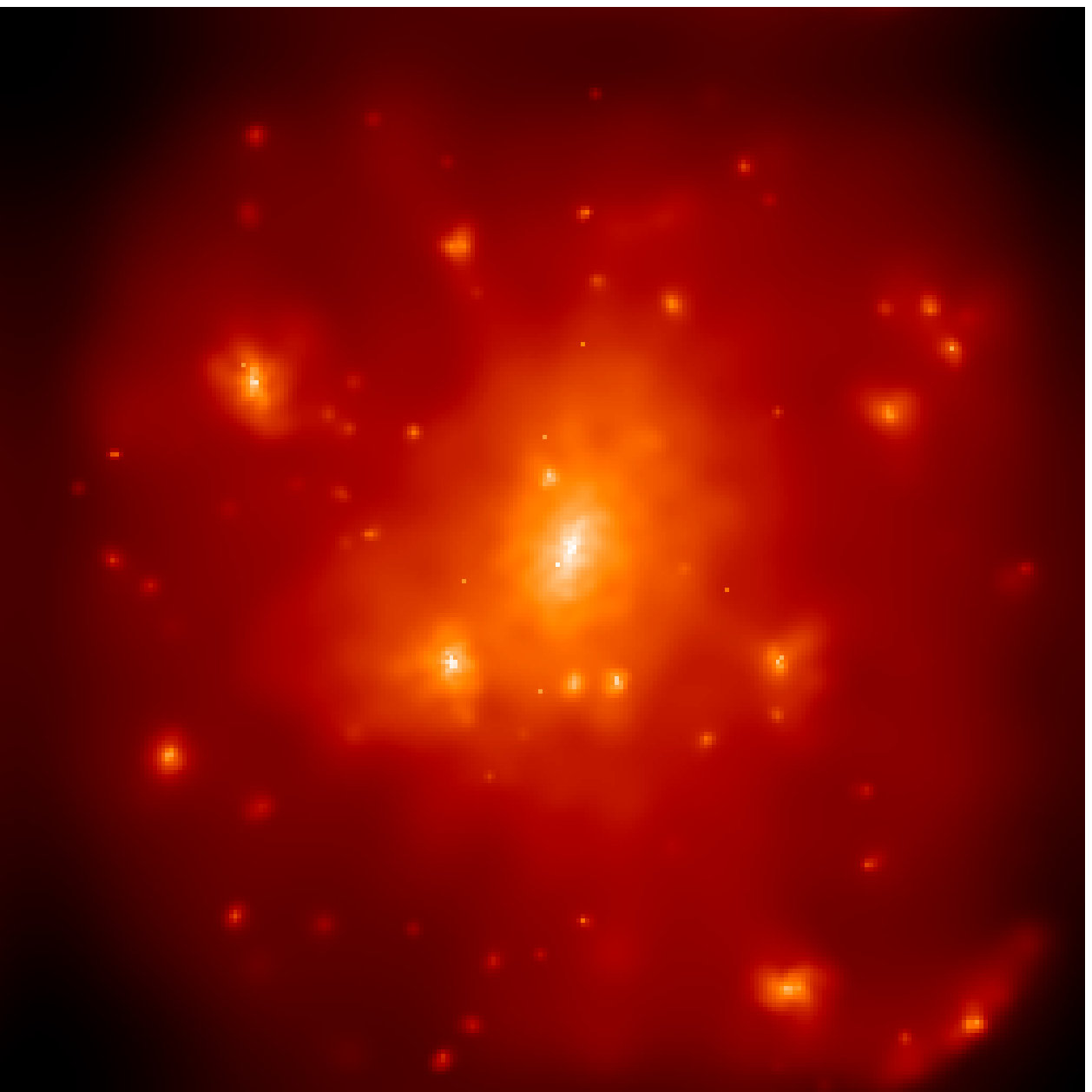}}
\hspace{-0.375cm}
\resizebox{3.64cm}{!}{\includegraphics{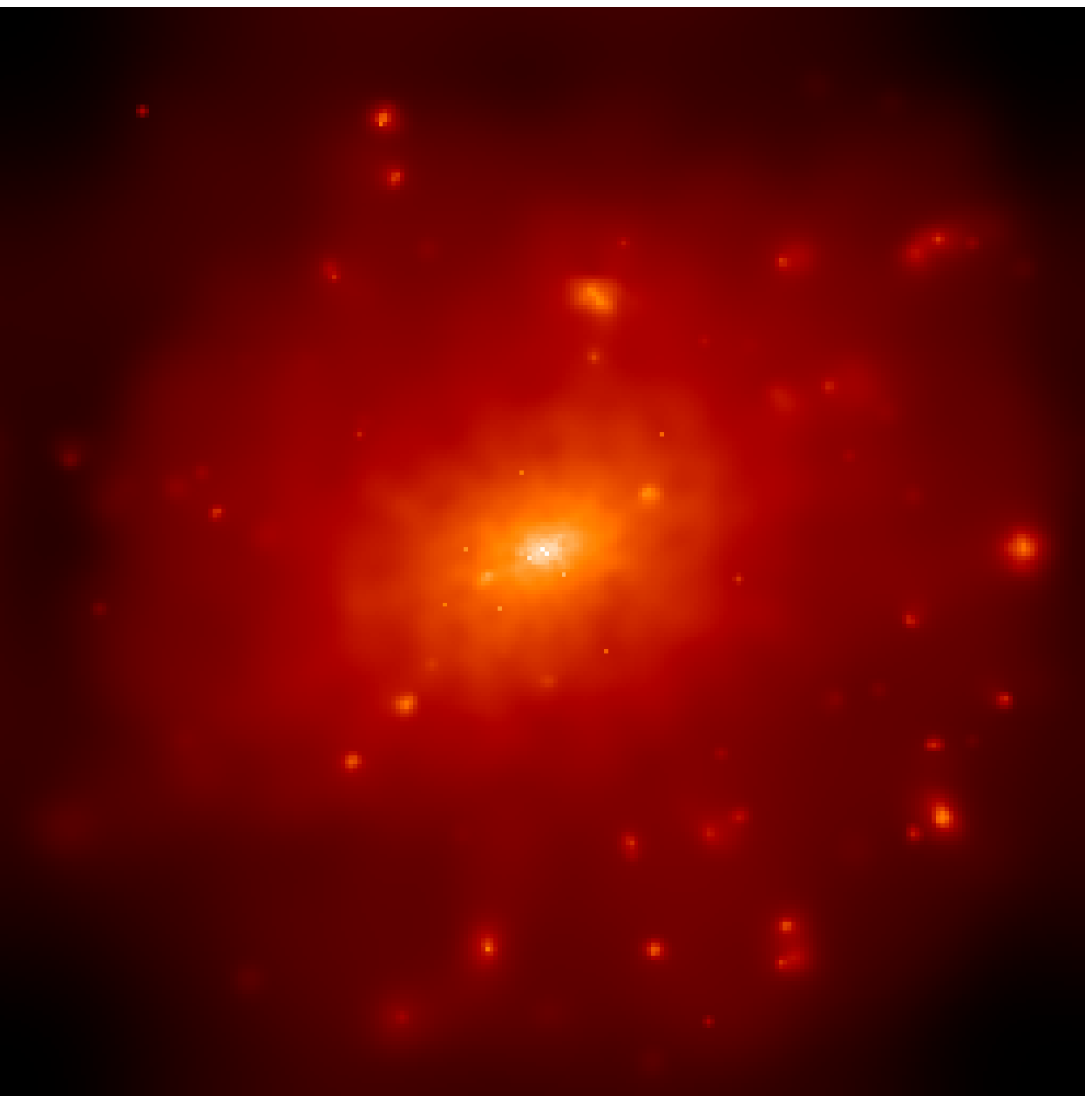}}
} 
\mbox{
\hspace{-0.6cm}
\resizebox{8.7cm}{!}{\includegraphics{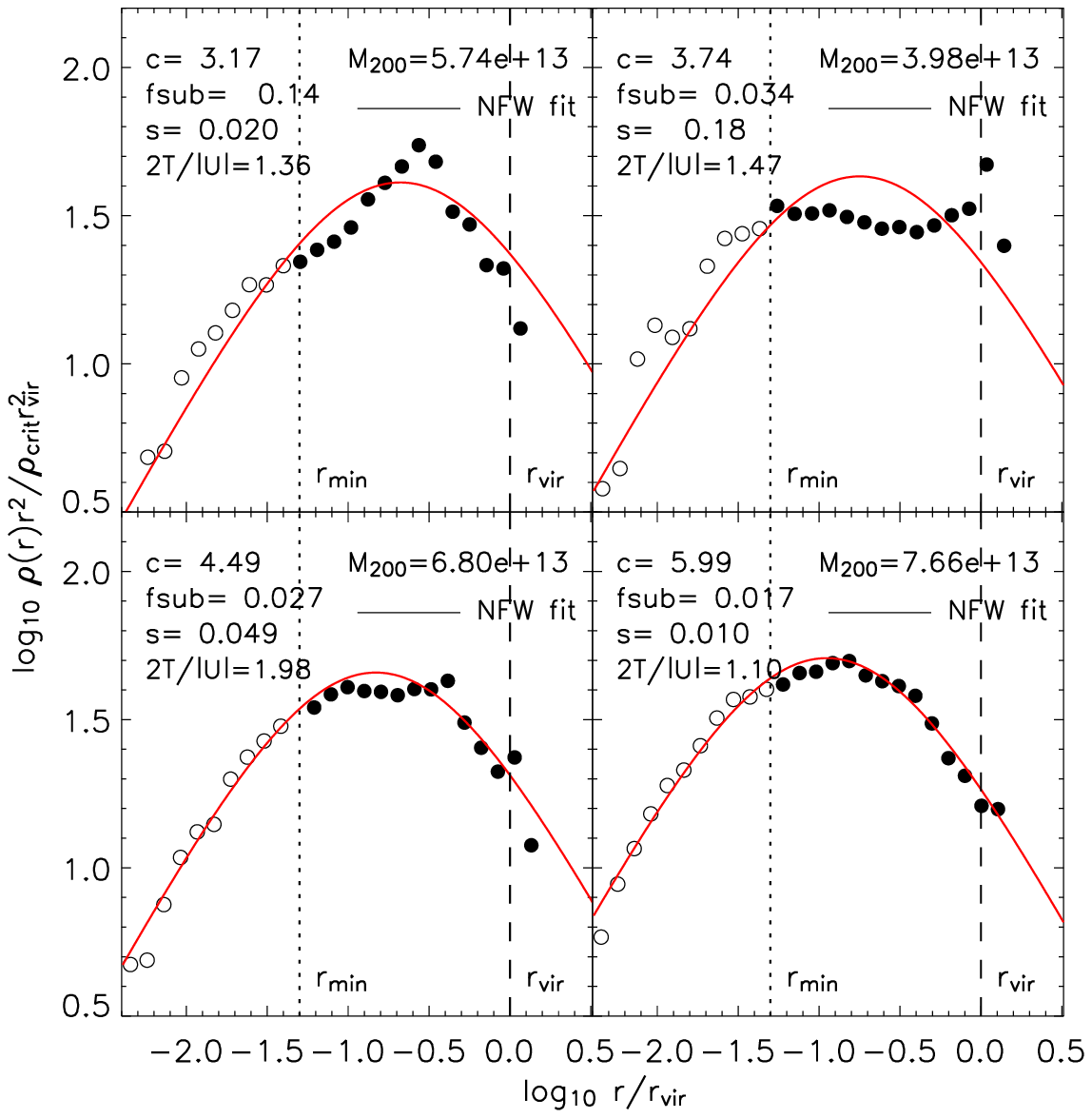}}
}
\end{center}
\caption{Images (top) and corresponding spherically averaged density
profiles (bottom) in four haloes of similar mass.  The halo shown in
the lower right panel of each set satisfies all our selection criteria
and is, therefore, close to dynamical equilibrium. Note that the NFW
profile (solid line) provides an excellent fit to this halo.  The halo
mass, concentration and the values of the three quantities, $f_{\rm
sub}$, $s$ and $2T/|U|$ used in the selection are given in the legend.
The NFW fitting procedure, which here is performed only over the
indicated range $r_{\rm min}<r<r_{\rm vir}$, is described in
Section~\ref{ssec:proffit}.  The remaining three haloes are excluded
from our {\it relaxed} sample as they fail at least one of the
selection criteria.  The halo on the upper left has a large amount of
substructure, $f_{\rm sub}>0.1$.  The one in the upper right panel is
undergoing a major merger. Note that the merging partner does not
contribute to $f_{\rm sub}$, since its centre lies outside the virial
radius, but some of its associated material displaces the centre of
mass of the system, resulting in $s>0.07$. The halo in the lower left
panel satisfies these two criteria, but has $2T/|U|>1.35$.  The
corresponding panels in the lower plot show that these unrelaxed haloes
have density profiles that are clearly not well described by NFW
profiles.}
\label{fig:pictures}
\end{figure}
%%%%%%%%%%%%%%%%%%%%%%%%%%%%%%%%%%%%%%%%%%%%%%

\subsection{Halo selection}
\label{ssec:hsel}

Dark matter haloes are dynamic structures, constantly accreting
material and often substantially out of virial equilibrium. In these
circumstances, haloes evolve quickly, so that the parameters used to
specify their properties change rapidly and are thus
ill-defined. Furthermore, in the case of an ongoing major merger, even
the definition of the halo centre becomes ambiguous, so that the
characterisation of a system by spherically-averaged profiles is of
little use. As we shall see below, departures from equilibrium not
only add to the scatter in the correlations that we seek to establish,
but can also bias the resulting trends, unless care is taken to
identify and correct for the effect of these transient structures.

\subsubsection{Relaxed and unrelaxed haloes}
\label{sssec:relaxcrit}

The equilibrium state of each halo is assessed by means of three
objective criteria:

\begin{itemize}

\item[i)]{\bf Substructure mass fraction:} We compute the mass
 fraction in resolved substructures whose centres lie inside $r_{\rm
 vir}$: $f_{\rm sub}=\sum_{i\neq0}^{N_{\rm sub}}M_{\rm sub,i}/M_{\rm
 vir}$.  Note that in this definition $f_{\rm sub}$ does not include
 the most massive substructure as this is simply the bound component
 of the main halo.

\item[ii)]{\bf Centre of mass displacement:} We define $s$, the
normalised offset between the centre of mass of the halo (computed
using all particles within $r_{\rm vir}$) and the potential centre, as
$s =|{\mathbf r}_{\rm c}-{\mathbf r}_{\rm cm}|/r_{\rm vir}$
\citep{thomas01}.

\item[iii)]{\bf Virial ratio:} We compute $2T/|U|$, where $T$ is the
total kinetic energy of the halo particles within $r_{\rm vir}$ and
$U$ their gravitational self potential energy.  To estimate $U$, we
use a random sample of $1000$ particles when $N_{\rm i}\geq 1000$. We
obtain physical velocities with respect to the potential centre by
adding the Hubble flow to the peculiar velocities and then we compute
the halo kinetic energy after subtracting from the velocities the
motion of the halo centre of mass.

\end{itemize}

Fig.~\ref{fig:pictures} shows images and spherically averaged density
profiles for a set of three haloes of similar mass that are rejected by
just one of each of the above criteria. Systems such as the one shown
in the top-left panel (large $f_{\rm sub}$) are clearly not relaxed
and it is not surprising that the large number of substructures
affect halo properties such as concentration, angular momentum and
shape \citep{gao04,shaw05}. Note that, despite the large
value of $f_{\rm sub}$, the centre of mass displacement is quite
small, since the spatial distribution of substructures happens to be
fairly symmetric.

Conversely, the halo shown in the upper-right panel of
Fig.~\ref{fig:pictures} has small $f_{\rm sub}$, but is rejected by
our cut on the centre of mass offset $s$. This is clearly an ongoing
merger where, because the merging partner of the main halo lies just
outside the virial radius, it contributes little to $f_{\rm sub}$.
Thus the $s$ criterion is complementary to $f_{\rm sub}$. This is
important, because the ability of SUBFIND to detect self-bound
substructures is heavily resolution-dependent: $f_{\rm sub}$ is likely
to be substantially underestimated in low-mass haloes resolved with few
particles.

These two criteria, however, make no use of kinematic information and
so our third criterion, based on $2T/|U|$, is a useful supplement able
to reject haloes that, despite passing the other two conditions, are
far from dynamical equilibrium. This is especially the case for
ongoing mergers and artificially linked haloes. An example of a halo
rejected by just this criterion is shown in the lower-left panel of
Fig.~\ref{fig:pictures}.

Finally, the lower-right panel of Fig.~\ref{fig:pictures} shows an
example of a halo that makes it into our {\it relaxed} sample.  These
relaxed haloes generally have smooth density profiles and are well
fitted by NFW profiles.

%%%%%%%%%%%%%%%%%%%%%%%%%%%%%%%%%%%%%%%%%
\begin{figure}
\mbox{
\hspace{-0.75cm}
\resizebox{9cm}{!}{\includegraphics{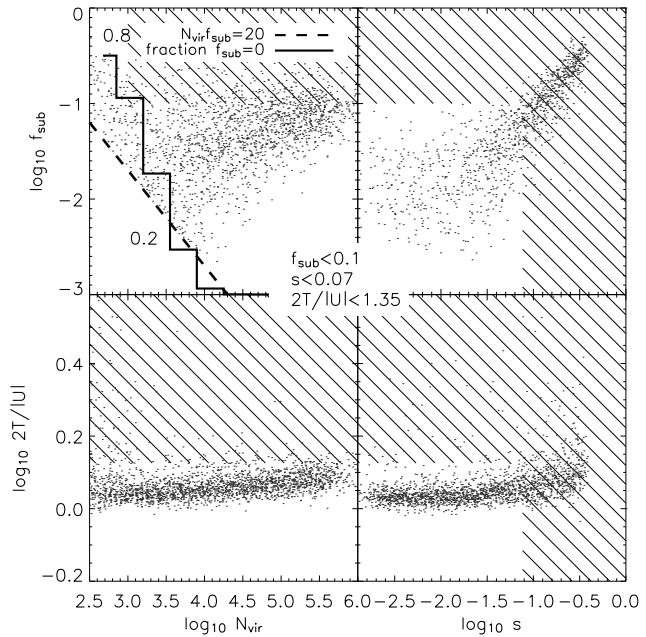}}
}
\caption{The various criteria used to define our {\it relaxed} sample
of haloes (see Section~\ref{ssec:hsel} for definitions), shown as a function
of the number of particles within the virial radius. Shaded regions
indicate the location of ``unrelaxed'' haloes. The dots sample
uniformly all {\sl MS} haloes with $N_{\rm vir}>300$ in each plot.
{\it Top left:} The fraction of halo mass in resolved substructures,
$f_{\rm sub}$ vs. the number of particles in the halo $N$. The dashed
line shows the detection limit of SUBFIND, $N_{\rm min}=20/N$. The
solid stepped line shows the fraction of haloes with no resolved
substructures in each mass bin (i.e., those with $f_{\rm sub}=0$).
{\it Top right:} The centre of mass offset $s$ vs the substructure
fraction $f_{\rm sub}$. The criteria $f_{\rm sub}<0.1$ and $s<0.07$
reject the tail of haloes with high $f_{\rm sub}$ and $s$.  {\it
Bottom:} The criterion $2T/|U|<1.35$ is intended to reject haloes far
from dynamical equilibrium.}
\label{fig:selection}
\end{figure}
%%%%%%%%%%%%%%%%%%%%%%%%%%%%%%%%%%%%%%%%%

\subsubsection{N-dependence of selection criteria}
\label{sssec:ndep}

Fig.~\ref{fig:selection} shows the various correlations between the
criteria used to identify ``relaxed'' haloes ($f_{\rm sub}$, $s$ and
$2T/|U|$) and the number of particles within the virial radius. Note
that the equilibrium measures we use do not have bimodal
distributions, but, instead, are roughly continuous, with extended
tails.  This reflects the fact that haloes assemble hierarchically: all
haloes are in the process of accreting some material and are,
therefore, to some extent unrelaxed. (Note for example, that the
median $2T/|U|$ is slightly greater than unity, as reported in earlier
work \cite[see, e.g.][]{cole96}). Thus, our selection criteria just
provide a simple, but somewhat arbitrary, way of trimming off the tail
of worst offenders, typically ongoing mergers.

The shaded regions in Fig.~\ref{fig:selection} show the areas of
parameter space rejected by our criteria to select {\it relaxed}
haloes: $f_{\rm sub}<0.1$, $s<0.07$ and $2T/|U|<1.35$.  The top-right
panel shows the expected strong correlation between our measures of
asymmetry and lumpiness, as well as how the tail of high $s$ and high
$f_{\rm sub}$ values is removed by the selection.

Of all the selection criteria, the one most sensitive to the number of
particles in the halo is $f_{\rm sub}$. This is clearly seen in the
top-left panel of Fig.~\ref{fig:selection}, where the stepped line
shows, as a function of $N_{\rm vir}$, the fraction of haloes with no resolved
substructures, i.e., $f_{\rm sub}=0$. This rises quickly with
decreasing $N_{\rm vir}$, so that more than $80\%$ of haloes with $300<N_{\rm vir}<1000$
particles have no discernible substructures. By contrast, essentially
{\it all} haloes with more than $10,000$ particles have at least one
massive subhalo within the virial radius.

A significant fraction of haloes with fewer than $1000$ particles also
have very large values of the virial ratio $2T/|U|$, as shown in the
bottom-left panel of Fig.~\ref{fig:selection}. As discussed by \citet{bett07}, 
these are loosely bound objects connected by a tenuous
bridge of particles, or objects lying in the periphery of much more
massive systems that owe their large kinetic energy to contamination
with fast-moving members of the other system.

These results suggest that a minimum number of particles may also be
required in order to obtain robust results independent of numerical
artifact. We shall see below that about $\sim 1000$ particles or more
are needed in order to avoid biases when deriving structural
parameters from fits to the halo mass profiles.

The fraction of haloes rejected by these cuts varies slowly with mass,
rising from 20\% at $M_{\rm vir} =10^{12}\, h^{-1} M_\odot$ to 50\% at
$M_{\rm vir} =10^{15}\, h^{-1} M_\odot$ (see numbers given in
Fig.~\ref{fig:massc} and Table~\ref{tab:fits}).  The criterion
$s<0.07$ rejects the most haloes; a smaller but significant fraction of
haloes that pass that cut are removed by the $f_{\rm sub}<0.1$
criterion; while only a few additional haloes are rejected by the
$2T/|U|<1.35$ test.

%We adopt $M>10^{13}h^{-1}$~M$_\odot$ for haloes selected from the
%Millennium Run and $M>10^{12}h^{-1}$~M$_\odot$ for haloes selected from
%the higher resolution simulation.
%Finally, as we
%wish to measure halo concentrations it is equally important that the
%resolution be sufficient that the measured concentration is well
%determined and robust.

\subsection{Merger trees and formation times}
\label{ssec:mergertrees}

We use the merger trees for the {\sl MS} described in detail by
\cite{harker06} and \cite{bower06}. These differ slightly from those
used by \cite{springel05b}, but the differences are only significant
for haloes undergoing major mergers, which would not pass the stringent
selection criteria listed above.  At each of the approximately $60$
output redshifts of the simulation, the merger tree provides us with a
list of all the haloes that will subsequently merge to become part of
the final halo. We exploit this information below in order to
investigate the dependence of halo concentration on formation
history (see Section~\ref{ssec:cpred}).

\begin{figure}
\epsfxsize=8.0truecm \epsfbox{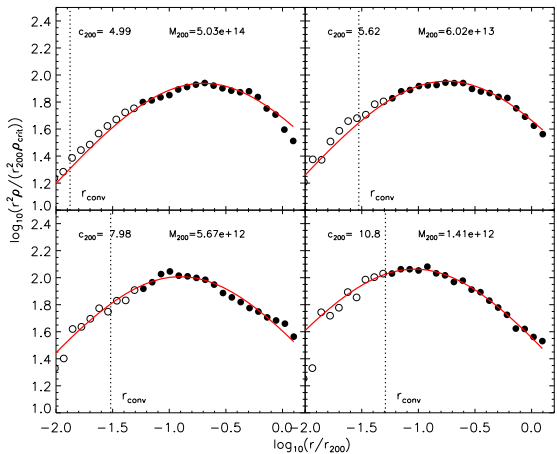} 
\caption{Density profiles, $r^2 \rho(r)$, and least squares NFW fits
for four {\it relaxed} haloes. The fits are performed over the radial
range $0.05 <r/r_{\rm vir}<1$, shown by the solid circles, and extend
slightly beyond $r_{200}$.  The vertical line marks the position of
the mass profile convergence radius derived by Power et al (2003). This
is always smaller than the minimum fit radius, $0.05 \, r_{\rm vir}$.}
\label{fig:profiles}
\end{figure}

\subsection{Profile fitting}
\label{ssec:proffit}

For each halo identified using the procedure outlined in
Section~\ref{ssec:hid} we have computed a spherically-averaged density
profile by binning the halo mass in equally spaced bins in
$\log_{10}(r)$, between the virial radius and $\log_{10}(r/r_{\rm
vir})=-2.5$. After extensive testing, we concluded that using 32 bins,
corresponding to $\Delta\log_{10}(r)=-0.078$, is enough to produce
robust and unbiased results.

The density profiles of four {\it relaxed} haloes are shown in
Fig.~\ref{fig:profiles}, together with fits using the NFW profile
(eq.~\ref{eq:nfw}). This profile has two free parameters, $\delta_c$
and $r_s$, which are both adjusted to minimise the rms deviation,
$\rms$, between the binned $\rho(r)$ and the NFW profile,
\begin{equation}
\rms^2 = \frac{1}{N_{\rm bins} -1}\sum_{i=1}^{N_{\rm bins}}
{\bigl[\log_{10}\rho_{\rm i} - \log_{10} \rho_{\rm NFW}(\delta_c;r_s)
\bigr]^2}.
\label{eq:rms}
\end{equation}
Note that eq.~\ref{eq:rms} assigns equal weight to each bin. We have
explicitly checked that none of our results varies significantly if we
adopt other plausible choices, such as Poisson weighting each bin (for
further discussion, see \cite{jing00}).

Once the parameters $\delta_c$ and $r_s$ for each halo are retrieved
from the fitting procedure, they may be expressed in a variety of
forms. In order to be consistent with the original NFW work, we choose
to express the results in terms of a mass, $M_{200}$, and a
concentration, $c=c_{200}=r_{200}/r_s$, that result from adopting
$\Delta=200$ in the definition of the virial properties of a halo
(eq.~\ref{eq:rvir}).

The radial range adopted for the fitting procedure is important,
especially since haloes of different mass are resolved to varying
degree in a single cosmological simulation. After experimenting at
length, and especially after comparing the fit parameters obtained in
the overlapping mass range of the two simulations (Section~\ref{ssec:ms}),
we settled on carrying out the fits over a uniform radial range (in
virial units). This ensures that all haloes, regardless of mass, are
treated equally, minimising the possibility of introducing subtle
mass-dependent biases in the analysis.

\begin{figure}
\epsfxsize=8.0truecm \epsfbox{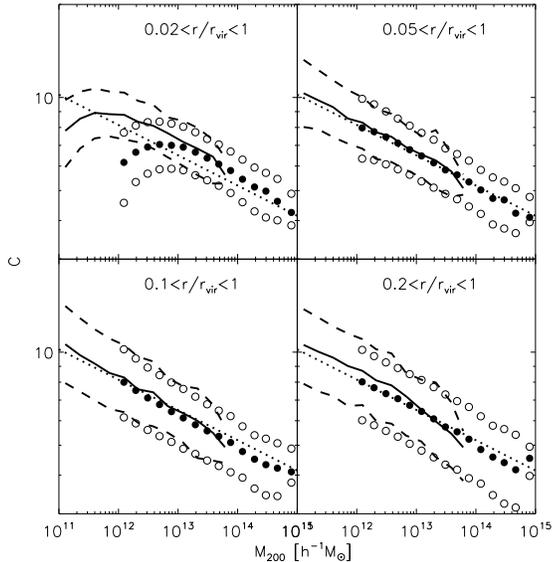} 
\caption{The median, $20$ and $80$ percentiles of the concentration,
$c_{200}$, as a function of halo mass, $M_{200}$. The symbols
extending to high masses show the results from the {\sl MS} while the
overlapping set of solid and dotted lines extending to lower masses
show the results from a simulation with $9 \times$ higher mass
resolution. Each panel corresponds to a different radial range adopted
for the fits. Data for each simulation are shown for haloes with
$N>1000$ particles, corresponding to $\sim 10^{12} h^{-1} \,
M_{\odot}$ in the {\sl MS}. The dotted line shows a power law,
$c\propto M^{-1/10}$, and is the same in all panels.}
\label{fig:resolution}
\end{figure}

Fig~\ref{fig:resolution} shows the mass-concentration dependence
obtained for the {\sl MS} (symbols) and the higher mass resolution
simulation (lines) for four different choices of the radial range. The
symbols and lines extend down to haloes with $\sim 1000$ particles
within the virial radius in each case and indicate the median (solid
circles and lines) as well as the 20 and 80 percentiles of the
distribution. To guide the comparison, the dotted line shows a power
law, $c\propto M_{200}^{-1/10}$, and is the same in all panels.

Note that, provided the minimum radius is $\ge 0.05\, r_{\rm vir}$,
the fit parameters appear to depend only weakly on the radial range,
but that the distribution of concentrations is narrowest for
$0.05<r/r_{\rm vir}<1$. For this choice (which we adopt hereafter in
our analysis) there is also good agreement between the two simulations
on mass scales probed in both with at least $1000$ particles.

\begin{figure} 
\includegraphics[width=9cm]{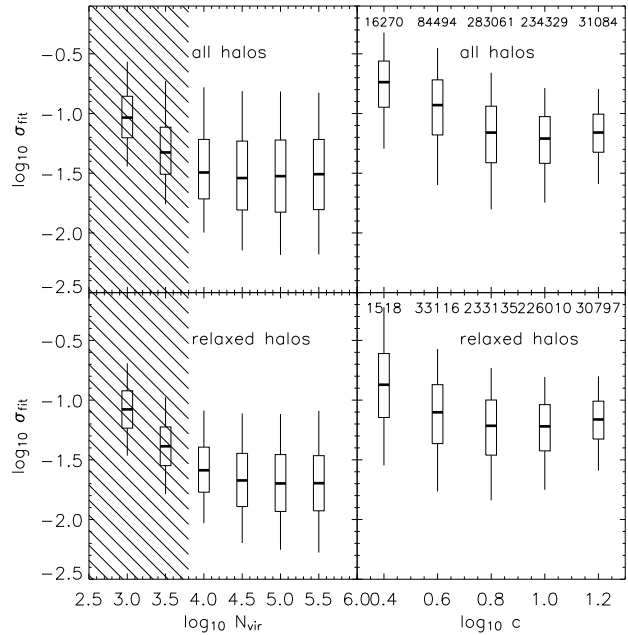}
\caption{The dependence of the rms residual deviation, $\sigma_{\rm
fit}$, about the best-fitting NFW density profile on the number of
particles per halo and on halo concentration. The boxes show the
medians and the $25\%$ and $75\%$ centiles of the distribution, while
the whiskers show the $5\%$ and $95\%$ tails. The numbers along the
top of each panel indicate the number of haloes within each bin.  Top
panels include all {\sl MS} haloes with $N_{\rm vir}>450$ as no other selection
criteria have been applied. The upturn in $\sigma_{\rm fit}$ for
low-concentration relaxed haloes is due to the inclusion of haloes with
less than $10,000$ particles. This upturn disappears once the
$N>10,000$ criterion is imposed.}
\label{fig:crms}
\end{figure}

However, as shown in Fig.~\ref{fig:crms}, the mean quality of the fits,
as measured by $\sigma_{\rm fit}$, deteriorates noticeably for haloes
with less than $10,000$ particles as a result of the relatively small 
number of particles per bin. Although this does not seem to
introduce a bias in the mass-concentration relation, we have decided
to retain, conservatively, only haloes with $N>10,000$ for our
analysis, and to combine the two simulations in order to probe the
halo mass range $10^{12}<M/h^{-1}M_{\odot}<10^{15}$.

Finally, we considered a couple of alternative methods for
characterising the halo concentration that do not assume an NFW
density profile, such as the ratio of maximum to virial circular
velocities, $V_{\rm max}/V_{\rm vir}$ \citep{gao04} or the ratio of
masses enclosing different overdensities, such as $M_{\rm
vir}/M_{1000}$ \citep{thomas01}. These measures of the concentration,
presumably because they use information on just two particular points
in the profile, lead to substantially larger scatter in the
correlations that we examine here, so we do not consider them further
in this paper.

%%%%%%%%%%%%%%%%%%%%%%%%%%%
%%%%%%%%%%%%%%%%%%%%%%%%%%%

%%%%%%%%%%%%%%%%%%%%%%%%%%%
\begin{figure}
\includegraphics[width=9cm]{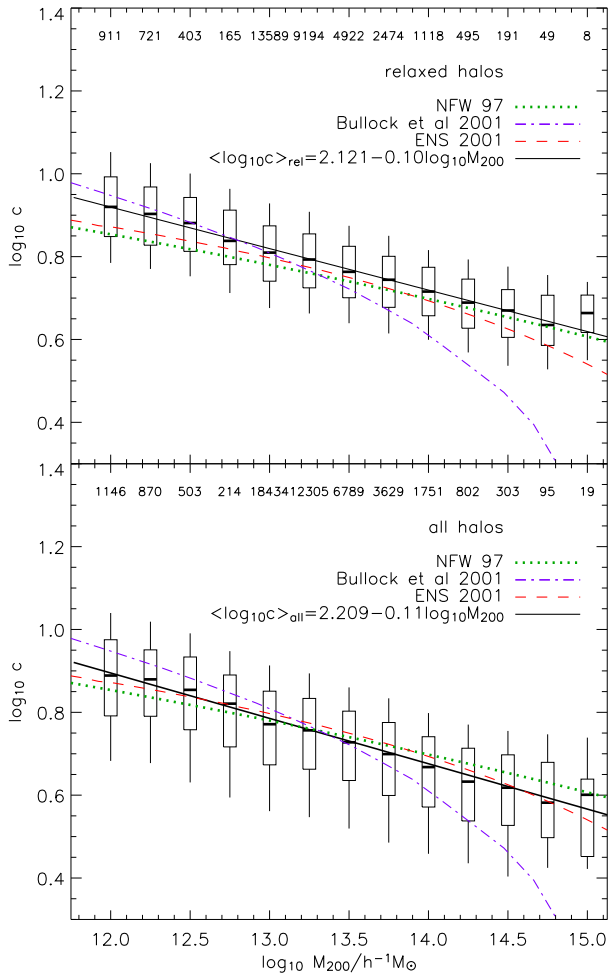}
\caption{Mass-concentration relation for {\it relaxed} haloes (top
panel) and for all haloes (bottom panel).  The boxes represent the
$25\%$ and $75\%$ centiles of the distribution, while the whiskers
show the $5\%$ and $95\%$ tails. The numbers on the top of each panel
indicate the number of haloes in each mass bin. The median
concentration as a function of mass is shown by the solid line and is
well fit by the linear relations given in the legend of each panel.
The dot-dashed line shows the prediction of the
\protect\cite{bullock01} model (using {\sl cvir2.f} available from the
authors); the dashed and dotted lines corresponds to the
\protect\cite{eke01} and NFW models, respectively, with
$\Gamma=0.15$, as this approximates best the input power spectrum of
the {\sl MS}. See text for further details.}
\label{fig:massc}
\end{figure}
%%%%%%%%%%%%%%%%%%%%%%%%%%%

\section{Results and Discussion}
\label{sec:res}

\subsection{Concentration vs mass}
\label{ssec:cvsm}

Fig.~\ref{fig:massc} shows concentration as a function of halo mass
for all haloes selected following the procedure described above. The
upper panel is for our {\it relaxed} halo sample, while the lower
panel show results for the complete sample, including systems that do
not meet our equilibrium criteria.

In both samples, the correlation between mass and concentration is
well defined, but rather weak. A power-law fits the median
concentration as a function of mass fairly well; we find:
\begin{equation}
c_{200} =  5.26 \, \left( 
M_{200}/10^{14} h^{-1} \, M_\odot \right)^{-0.10} ,
\label{eq:cmrel}
\end{equation}
for {\it relaxed} haloes, and 
\begin{equation}
  c_{200} =  4.67 \, \left( 
M_{200}/10^{14} h^{-1}\, M_\odot \right)^{-0.11} 
\label{eq:cmall}
\end{equation}
for the complete halo sample. 

Our power-law fit is in good agreement with the results of M07, 
who find $c_{200}\sim 5.6\, (M_{200}/10^{14} h^{-1} \,
M_\odot)^{-0.098}$ for the {\it average} concentration of their sample
of relaxed haloes. These authors also report that concentrations are
systematically lower when considering the full sample of haloes. The
small difference between our results and M07's may be due
to variations between mean and median, as well as on the different
criteria used to construct the relaxed halo sample. Nevertheless, the
agreement in the exponent of the power-law is remarkable, especially
considering that these authors explore a mass range different from
ours, namely $2\times 10^9 < M_{200}/h^{-1} \, M_\odot<2\times
10^{13}$. Combining these results with ours, we conclude that {\it a
single power law fits the concentration-mass dependence for about six
decades in mass.}

Over the mass range covered by our simulations the concentration-mass
dependence is in reasonable agreement with the predictions of NFW and
of ENS, as shown, respectively, by the dotted and dashed lines in
Fig.~\ref{fig:massc}
\footnote{These predictions use the original parameters in those
papers and have not been adjusted further, except for adopting the
power-spectrum ``shape'' parameter, $\Gamma=0.15$, as the best match
to the power spectrum adopted for the {\sl MS}.}
. The agreement, however, is not perfect, and both models appear to
underestimate somewhat the median concentration at the low mass end.

At the high-mass end, where there is a hint that concentrations are
approaching a constant value, the NFW model does slightly better than
ENS. This is because a constant concentration for very massive objects
is implicit in the NFW model, but not in ENS nor in the model of B01,
which is shown by a dot-dashed line in Fig.~\ref{fig:massc}. Both
ENS and B01 predict a strong decline in concentration at the very high
mass end. For the parameters favoured by B01, the disagreement for
$M>10^{13.5} h^{-1} M_\odot$ is dramatic, and cautions, as already
pointed out by \cite{zhao03b}, against using this model for
predicting the concentrations of massive haloes.

Finally, we note that M07 argue that the B01 model
reproduces their results better than ENS for haloes of mass a few times
$10^9 h^{-1} \, M_\odot$. However, the differences between the two
models only become appreciable below $\sim 10^{10} h^{-1} M_\odot$,
which corresponds to only about $700$ particles in their
highest-resolution simulation. Given (i) the large scatter in the
correlation; (ii) our finding that at least $1000$ particles are
needed to produce unbiased results, and (iii) the worries expressed in
Section~\ref{sec:intro} about the rather small box used by M07 
to resolve low-mass haloes, we conclude that it would be premature
to judge conclusively on the superiority of one model over the other
at the low mass end. More definitive tests are likely to come either
from much higher resolution simulations or from extending this
analysis to higher redshifts, where the two models predict different
behaviour. This is a topic that we intend to address in a forthcoming
paper (Gao et al. 2007, in preparation).

\subsubsection{Scatter}
\label{sssec:cmscat}

The large volume covered by the {\sl MS} means that we have a large
number of haloes in our sample, even in the most massive bins of
Fig.~\ref{fig:massc}, corresponding to rare, very massive objects
with masses approaching $10^{15} h^{-1} \, M_\odot$. We use this to
check the common assumption that, at given halo mass, concentrations
are distributed following a lognormal distribution \citep{jing00}. 

\begin{figure}
\includegraphics[width=9cm]{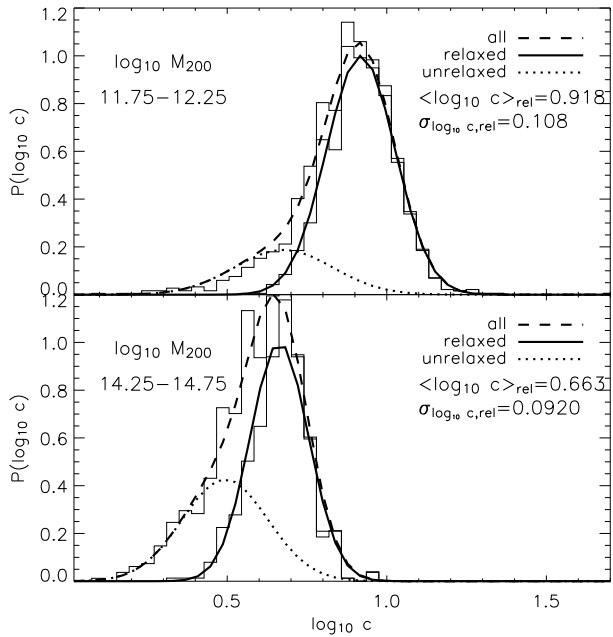}
\caption{The distribution of concentrations for haloes in two different
mass bins: $11.75< \log_{10} M_{200}/h^{-1}\, M_\odot<12.25$ (top
panel) and $14.25<\log_{10} M_{200}/h^{-1}\, M_\odot <14.75$ (bottom
panel). The top histogram in each panel shows the distribution for all
haloes, the lower histogram that corresponding to the {\it relaxed}
halo sample. The smooth curves are lognormal fits.  Note that the
overall distribution may be very well approximated by the sum of two
lognormal functions with parameters listed in Table~\ref{tab:fits}.}
\label{fig:scatter}
\end{figure}

We examine this in Fig.~\ref{fig:scatter}, where we show the
distribution of concentrations in two mass bins. In each panel the top
histogram corresponds to {\it all} haloes in our sample, and the lower
histogram only to those deemed {\it relaxed} by the criteria listed in
Section~\ref{ssec:hsel}.

Although the overall distribution is {\it not} approximated well by a
lognormal function, that of {\it relaxed} haloes is. The thick lines in
Fig.~\ref{fig:scatter} show fits of the form
\begin{equation}
P(\log_{10}c)=\frac{1}{\sigma
\sqrt{2\pi}}\exp{\Bigl[-\frac{1}{2}\Bigl(\frac{\log_{10}c-
\langle\log_{10}c\rangle}
{\sigma}\Bigr)^2} \Bigr], 
\label{eq:lognormal}
\end{equation}
which is clearly a good approximation for the concentration
distribution of relaxed haloes.

Intriguingly, the concentration distribution of ``unrelaxed'' haloes
may also be approximated by a lognormal distribution, albeit of
smaller mean and larger dispersion, as shown by the dotted curves in
Fig.~\ref{fig:scatter}. The sum of the two lognormals (shown with
dashed lines) is indeed a very good approximation to the overall
distribution. We list the median concentration and dispersion of these
distributions, as well as the fraction of ``unrelaxed'' haloes, as a function of mass in Table~\ref{tab:fits}.

\begin{figure}
\includegraphics[width=9cm]{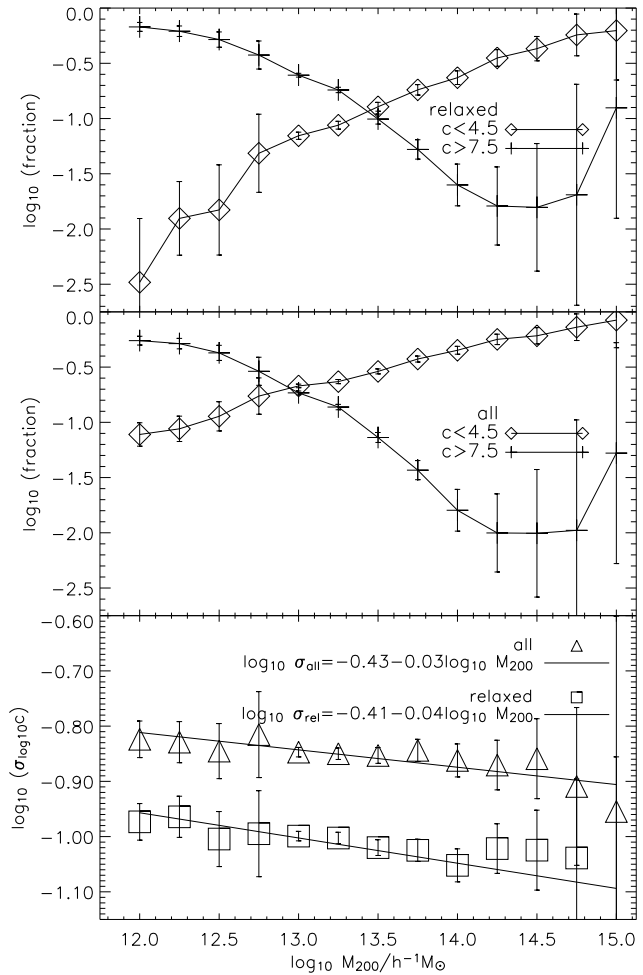}
\caption{The fraction of {\it relaxed} haloes (top panel) and {\it all}
haloes (middle panel) with $c<7.5$ (plus sign symbols) and $c>4.5$
(diamond-like symbols). The points connected by solid lines are the
direct measurements, in mass bins.  Error bars denote $1/\sqrt{N}$
uncertainties. The lower panel shows the measured rms scatter in
$\log_{10} c$ for {\it all} and {\it relaxed} haloes, respectively.}
\label{fig:fracscatter}
\end{figure}

Upon inspection, a weak but systematic trend is apparent in
Table~\ref{tab:fits}: the {\it dispersion} in concentration appears to
decrease monotonically as a function of mass (see also the bottom
panel of Fig~\ref{fig:fracscatter}). This suggests that massive haloes
are somehow a more homogeneous population than their lower mass
counterparts, and may reflect the fact that massive haloes are rare
objects that have collapsed recently, whereas less massive systems
have a much wider distribution of assembly redshifts. We shall return
to this topic below.

One may apply the distributions shown in Fig~\ref{fig:scatter} to
estimate the abundance of relatively rare objects. These might be,
for example, massive cluster haloes of unusually {\it high}
concentration that stand out in X-ray or lensing surveys, or perhaps
galaxy-sized haloes of unusually {\it low} concentration which may be
notable because of peculiarities in the rotation curves of the
galaxies they may host.

We show this in the top and middle panels of
Fig.~\ref{fig:fracscatter}, where the ``plus'' signs denote the
fraction of haloes in each mass bin that have $c<4.5$. The diamond
symbols, on the other hand, indicate the fraction of systems with
concentrations {\it exceeding} $7.5$. For example, we find that
slightly more than $1\%$ of {\it relaxed} haloes with $M_{200}\sim
3\times 10^{14} h^{-1}\, M_\odot$ have $c>7.5$, or about $4$ objects
in our $500^3 h^{-3}$ Mpc$^{3}$ volume. So clusters such as Abell
1689, for which \cite{limousin06} estimate $c_{200}=7.6$ should
not be very abundant in the local Universe.  Finding, through lensing
or X-ray studies, many clusters as massive and as concentrated as
Abell 1689 would yield strong constraints on the viability of
$\Lambda$CDM as a cosmological model.

%%%%%%%%%%%%%%%%%%%%%%%%%%%%%%%
%%%%%%%%%%%%%%%%%%%%%%%%%%%%%%%

%%%%%%%%%%%%%%%%%%%%%%%%%%%%%%%
%%%%%%%%%%%%%%%%%%%%%%%%%%%%%%%

\subsection{Concentration vs spin}
\label{ssec:cvsl}

Although models such as those proposed by NFW, B01 or ENS may more or
less account for the weak correlation between mass and concentration,
they say little about the origin of the sizable scatter about the mean
trend. As shown in Fig.~\ref{fig:scatter}, haloes of given mass have
concentrations that may differ by up to a factor of three or
more. What is the origin of this large scatter?

One possibility is that concentration may be related to the angular
momentum of the halo and that the large scatter in the spin parameter
at given mass (see, e.g., \cite{bett07} and references therein) is
somehow related to the spread in concentration. This was explored by
NFW and B01, who concluded that there was no obvious correlation
between spin and concentration; however, \cite{bailin05}  have
recently argued that such correlation indeed exists, and speculate
that this may explain why low surface-brightness galaxies (LSBs) appear
to inhabit low-density haloes \citep{deblok03,mcgaugh03}.

\begin{figure}
\includegraphics[width=14cm]{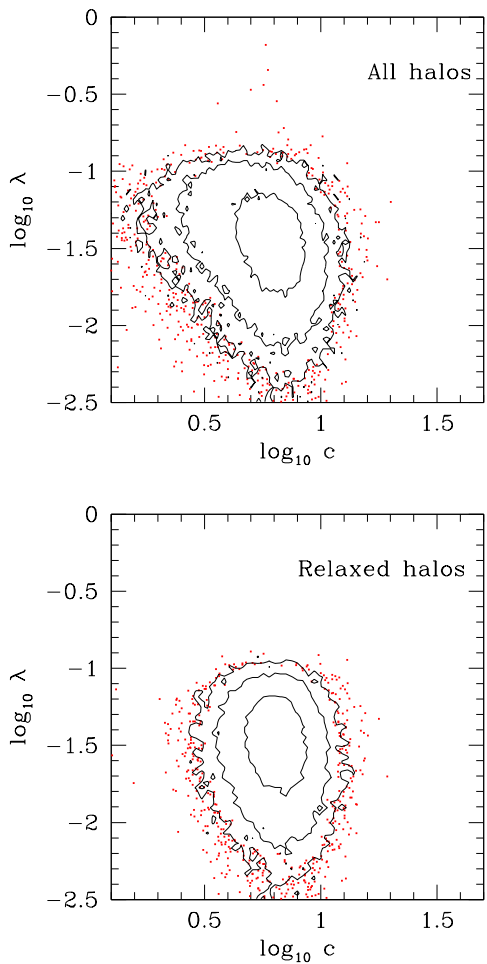}
\caption{The dependence of the spin parameter on concentration for the
full halo sample (top panel) and for {\it relaxed} haloes (bottom panel).  The
iso-density contours enclose 65, 95 and 99\% of the distribution while
the individual points show the distribution for the remaining 1\% of
the distribution.  }
\label{fig:cspin}
\end{figure}

We revisit this issue in Fig.~\ref{fig:cspin}, where we show the
spin parameter $\lambda=J|E|^{1/2}/GM^{5/2}$ (where $J$ is the angular
momentum with respect the potential centre and $E$ is the binding
energy of the halo) versus concentration for our full halo sample (top
panel) and for {\it relaxed} haloes (bottom panel). Although the full
halo sample shows a well-defined tail of low-concentration,
fast-spinning haloes reminiscent of \citeauthor{bailin05}'s claim, this
essentially disappears when {\it relaxed} haloes are considered.

As discussed above, the low concentrations of unrelaxed haloes measured
by profile-fitting algorithms are a transient result of the rapidly
evolving mass distribution that accompanies an accretion event, and do
{\it not} indicate a halo with lower-than-average density. 

However, this observation does not explain why, as is apparent from
Fig.~\ref{fig:cspin}, unrelaxed systems also tend to have, on
average, higher spins than their relaxed counterparts. As argued
recently by \cite{donghia07}, this is likely due to the
redistribution of mass and angular momentum that occurs during
mergers, combined with the arbitrary virial boundaries used to define
a halo and to compute its spin. Accretion events bring mass and
angular momentum into a system, but redistribute it in such a way that
high-angular momentum material ends up preferentially on weakly bound
orbits that may take it beyond the nominal virial radius, thereby
reducing $\lambda$. This effect seems to be particularly important
during mergers, and leads to an overall reduction of $\lambda$ as the
merger progresses (see also \cite{gardner01,vitvitska02}).

We conclude, in agreement with M07, that the bulk of
the effect reported by \cite{bailin05}  is due to the inclusion of
out-of-equilibrium haloes in their sample. A very weak correlation
between $c$ and $\lambda$ may be visible in the tilt of the contours
in the {\it relaxed} halo panel of Fig.~\ref{fig:cspin}, but it seems
too weak to have strong observational implications.

%%%%%%%%%%%%%%%%%%%%%%%%%%%%%%%
%%%%%%%%%%%%%%%%%%%%%%%%%%%%%%%

\subsection{Concentration vs formation time}
\label{ssec:cvszf}

The mass-concentration dependence was originally interpreted by NFW as
the result of the dependence of halo formation time on mass: the halo
characteristic density just reflects the mean density of the Universe
at the time of collapse. Therefore, low-mass haloes are more
concentrated because they collapsed earlier, when the Universe was
denser. NFW also showed that the bulk of the scatter in the
concentration at given mass is due to variations in formation
redshift.  In this section we revisit the NFW analysis using our
simulations.

The first thing to note is that there is no unique way of defining
when a particular halo formed, and so different definitions have been
adopted \citep{lacey93,nfw97,wechsler02,zhao03a}. The simplest and most
widely used definition is to set the formation redshift of a halo as
the time when the most massive progenitor first exceeds half of the
final halo mass, $M(z_f)=M(z=0)/2$.  We use this as our default
definition, but we will also consider other variants below.

\begin{figure}
\mbox{
\hspace{-0.75cm}
\resizebox{9cm}{!}{\includegraphics{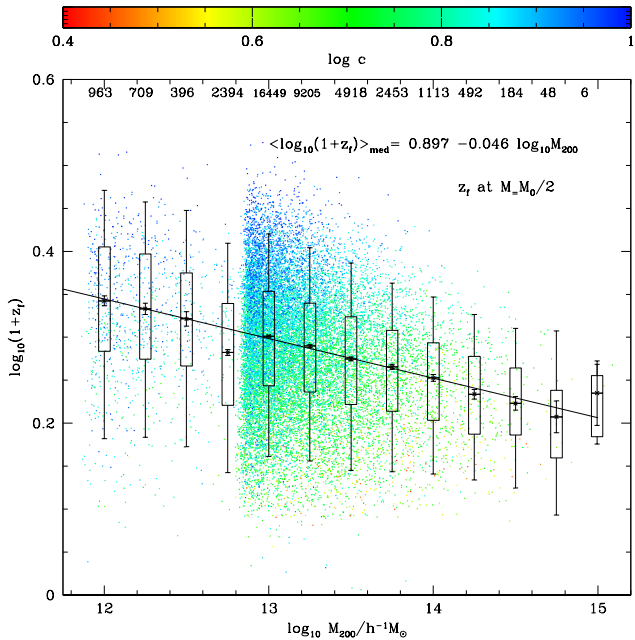}}
}
\caption{Halo formation time as a function of mass for {\it relaxed}
haloes. Formation times are defined as the time when the most massive
progenitor reaches half of the final halo mass. The numbers along the
top of the main panel indicate the number of haloes in each bin.  The
straight line is a least square fit to the median concentration. The
colour of the plotted points encodes the concentration as indicated by
the colour bar.  The gradual change in colour, from green in the upper
region of the plot to red at the bottom, shows qualitatively that
concentration depends sensitively on the formation time.  See also
Fig.~\ref{fig:residuals_zf}.  }
\label{fig:zf_mass}
\end{figure}

Fig.~\ref{fig:zf_mass} shows the dependence of the formation redshift
(expressed as $1+z_{\rm f}$) on halo mass. As expected, the median
formation redshift clearly declines with increasing mass, and the
solid line in the figure shows a least-squares fit to the median
concentration. Dots show a random selection of haloes coloured according
to their concentration. The gradual change in colour from top to bottom
indicates that concentration and formation time are closely related.

\begin{figure}
\includegraphics[width=9cm]{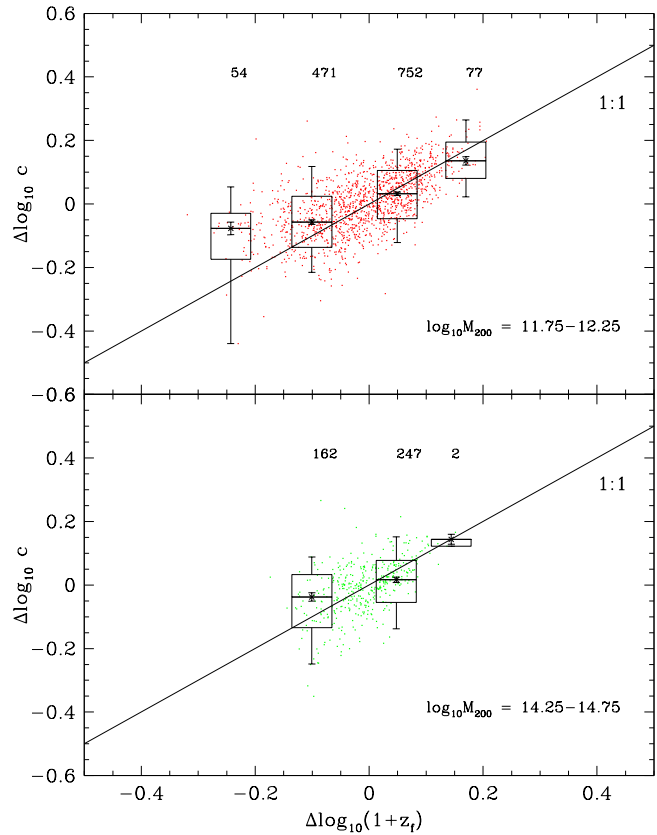}
\caption{Concentration offsets from the mean mass-concentration
relation (Fig.~\ref{fig:massc}) versus formation-time offsets from the
mean mass-formation time relation (Fig.~\ref{fig:zf_mass}) for two
mass bins, as labelled in each panel. The strong correlation between
residuals implies that much of the scatter in the $M$-$c$ relation is
due to variations in the formation time of haloes of given mass. Boxes
indicate the median and (25,75) percentiles, whiskers stretch to the
(5,95) percentiles. The number of haloes in each $\Delta \log_{10} (1+z_f)$ bin is given in the
legend.}
\label{fig:residuals_zf}
\end{figure}

We investigate this further in Fig~\ref{fig:residuals_zf}, where we
plot the offset in concentration from the mean mass-concentration
relation (Fig.~\ref{fig:massc}) versus the formation redshift offset
from the mean mass-formation redshift relation
(Fig.~\ref{fig:zf_mass}) for two mass bins. The strong correlation
between residuals indicates that the scatter in concentration at fixed
mass is mostly due to variations in formation time.

The fraction of the scatter in concentration shown in
Fig.~\ref{fig:scatter}, $\sigma_{\rm lgM}$, accounted for by $z_f$
variations may be estimated by comparing it with the rms scatter,
$\sigma_{\rm lgzf}$, about the 1:1 line in
Fig.~\ref{fig:residuals_zf}.  The fractional reduction in the
variance, $|\sigma_{\rm lgM}^2-\sigma_{\rm lgzf}^2|/\sigma_{\rm
lgM}^2$ is $35\%$ for the lowest mass range,
$10^{11.75}<M_{200}/h^{-1} M_\odot < 10^{12.25}$, and $12\%$ for the
highest mass range, $10^{14.25}<M_{200}/h^{-1} M_\odot < 10^{14.75}$.

This implies that the scatter in concentration is not fully explained
by differences in formation time alone, and that additional effects are
 at work.  One possibility is that the additional scatter is an
environmental effect, as recently proposed by \cite{wechsler06}.
However, these authors find that the effect is restricted to low-mass
haloes, whereas our results show additional scatter for high mass haloes
as well. 

Another possibility is that our formation time definition should be
revised. After all, the fraction of halo mass enclosed within the
scale radius, $r_s$ (which defines $c$), is {\it much less} than
one-half of the halo mass for the typical concentrations shown in
Fig.~\ref{fig:massc}. Or, finally, it might be that the scatter is
driven by some aspect of the merger history that is not fully captured
by our default definition of $z_{\rm f}$, which depends on a single
component of the halo (its most massive progenitor) rather than on the
full spectrum of fragments that coalesce to form the final halo.

\begin{figure}
\mbox{
\hspace{-0.75cm}
\resizebox{9cm}{!}{\includegraphics{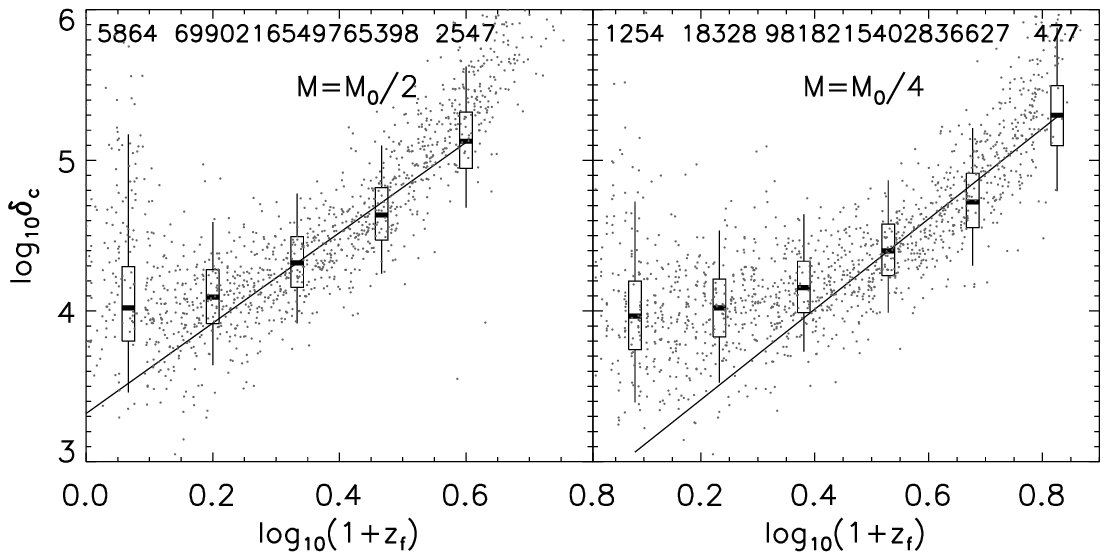}}
}
\mbox{
\hspace{-0.75cm}
\resizebox{9cm}{!}{\includegraphics{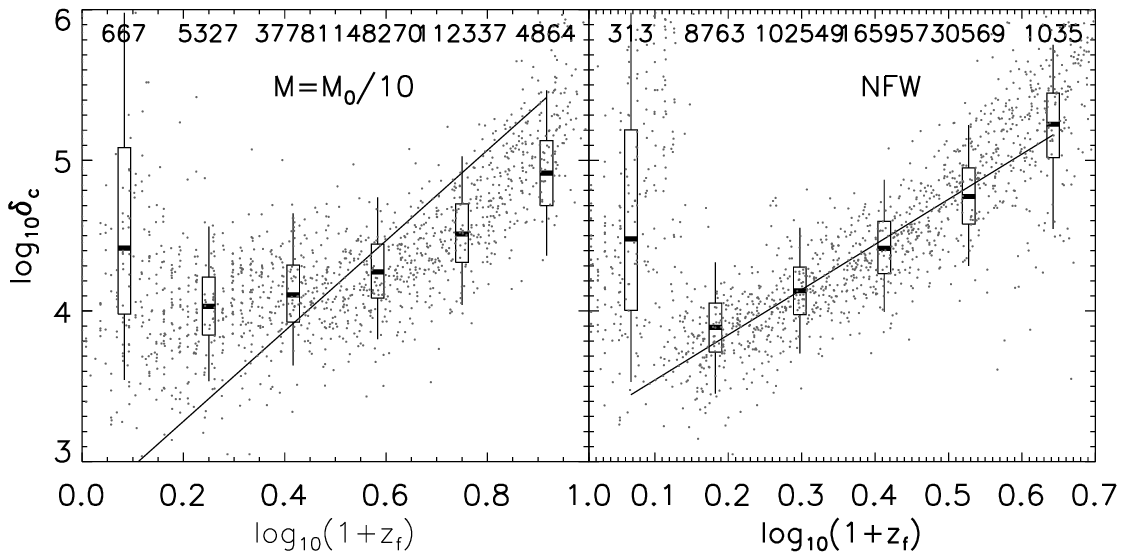}}
}
\caption{The correlation between the halo characteristic density,
 $\delta_c$, and different definitions of formation time for {\it
 relaxed} haloes. The numbers along the top of each panel indicate the
 number of haloes in each bin.  In the first three panels the formation
 time is defined by reference to the most massive progenitor, and is
 set to be the redshift at which its mass was $1/2$, $1/4$ or $1/10$
 of the final halo mass, $M_0$ (see labels in each panel). In the
 final panel (labelled NFW) the formation time is defined as the
 redshift when half of the final halo mass is in progenitors more
 massive than $1/10$ of $M_0$.  The ``natural scaling"
 $\delta_c\propto(1+z_f)^3$ is indicated by the solid line in each
 panel.}
 \label{fig:scatterzf}
\end{figure}

We explore this in Fig.~\ref{fig:scatterzf}, where we compare the
NFW characteristic density, $\delta_c$, with four different
definitions of the formation redshift. We use $\delta_c$ in this
figure because, besides being equivalent to the concentration, $c$, it
is easier to interpret. Indeed, if the
characteristic density really tracks the mean density of the universe
at the time of formation, one would expect it to follow the ``natural''
scaling, $\delta_c \propto (1+z_{\rm f})^3$.

Three of the $z_f$ definitions explored in Fig.~\ref{fig:scatterzf}
are variations of the one adopted above, and indicate the time when
the most massive progenitor first exceeds 10, 25, and 50\% of the
final halo mass, $M_0=M(z=0)$. These correspond to the panels labelled
$M_0/10$, $M_0/4$, and $M_0/2$ in Fig.~\ref{fig:scatterzf},
respectively. The last definition, on the other hand, follows the
prescription of NFW, and identifies the time when the {\it combined
mass} of all $M>M_0/10$ progenitors exceeds $M_0/2$.

As is clear from Fig.~\ref{fig:scatterzf}, the tightest relation
around the ``natural scaling'' (shown as solid lines) corresponds to
the NFW definition.  This suggests that the full mass spectrum of
clumps that assembles into the halo plays an important role in the final
halo concentration and not just the most massive progenitor.

We also note that in each panel of Fig.~\ref{fig:scatterzf} it is the
first (lowest redshift) bin that has the largest scatter and that is
furthest form the ``natural scaling" relation.  This bin corresponds
to haloes that have been assembled very recently, and may contain haloes
that, despite our selection criteria, are still unrelaxed and for
which the structural parameters are ill defined.

%%%%%%%%%%%%%%%%%%%%%%%%%%%%%%%%%%%%
%%%%%%%%%%%%%%%%%%%%%%%%%%%%%%%%%%%%

%%%%%%%%%%%%%%%%%%%%%%%%%%%%%%%%%%%
%%%%%%%%%%%%%%%%%%%%%%%%%%%%%%%%%%%

%%%%%%%%%%%%%%%%%%%%%%%%%%%%%%%%%%%%%
%%%%%%%%%%%%%%%%%%%%%%%%%%%%%%%%%%%%%

\subsection{Concentration predictions}
\label{ssec:cpred}

It is clear from the above discussion that accurate predictions of the
concentration require some knowledge of the halo's assembly
history. This cautions against the common practise in semi-analytic
models of assigning
concentrations to haloes according to just their mass and to some
probabilistic accounting of the dispersion shown in
Fig.~\ref{fig:scatter}. As emphasised by \cite{gao05},
properties such as the clustering of haloes depend on the assembly
history, so a full description of the correlation between formation
history and concentration may affect significantly the model
predictions for the size and internal structure of a galaxy.

% This in
%turn could have important implications for the predictions of the
%Tully-Fisher relation of gas rich, disc dominated galaxies or for the
%properties of unusually low surface brightness dwarfs.

A couple of prescriptions designed to predict halo concentrations at
$z=0$ from their mass accretion histories have been proposed recently
(W02, Z03), and we use our simulation merger trees in
order to compare them. We focus on the {\it relaxed} halo sample,
since these have well-defined concentration parameters.  We also
compare these methods with the simple prescription originally proposed
by NFW.

\subsubsection{Wechsler et al. prescription}
\label{sssec:wechsler}

W02 showed that the Mass Accretion History (MAH) of a
halo's most massive progenitor, of mass $M_0$ at redshift $z=0$, may
be approximated by a simple function,
\begin{equation}
\log_{10} M(z)=\log_{10} M_0 - \alpha z
\label{eq:wechsler}
\end{equation}
i.e., by a straight line with slope $-\alpha$ in the plane
$\log_{10}M(z)$ vs. $z$ (see also \cite{vandenbosch02}). These authors
then relate the parameter $\alpha$ to a formation time via $a_{\rm
f}=1/(1+z_{\rm f})=\alpha/2\ln(10)$.  Their Fig.~6 shows that this
definition of formation time correlates well with the halo
concentrations measured by B01 in their simulations and
can be used to predict $c$ at $z=0$.

Fitting this correlation, they find
\begin{equation}
c_{\rm W}=c_{\rm 0}/a_{\rm f},
\label{eq:wcal}
\end{equation}
with $c_{\rm 0}=4.1$, the typical concentration of haloes forming at
the present time.  The implementation of this prescription in our
simulations is straightforward, and we show the predictions in
Fig.~\ref{fig:predictions}, after recalibrating eq.~\ref{eq:wcal} with
$c_{\rm 0}=2.26$ in order to take into account our different
definition of virial radius.

\subsubsection{Zhao et al. prescription}
\label{sssec:zhao}

Z03 differentiate two distinct phases in the MAH; one of
early, fast accretion, followed by a slow-accretion period that lasts
until the present. The transition between the two phases occurs at a
characteristic redshift, $z_{\rm tp}$. This ``turning-point'' redshift
may be used to estimate the concentration, assuming that the inner
properties of the halo, such as the scale radius, $r_s$, and its
enclosed mass, are set at $z_{\rm tp}$ and vary weakly thereafter.

This procedure is in principle straightforward to implement in our
simulations but we note that there are a substantial number of haloes
for which the distinction between the two accretion phases is not
well-defined. In some cases, more than one phase of fast accretion
seems to be present; in others, there is a single phase with no
obvious turning point.  This leads to ambiguities in the definition of
$z_{\rm tp}$ and its associated concentration that are not easily
resolved and that affect a significant fraction of haloes. A similar
worry applies to the Wechsler et al prescription, since
eq.~\ref{eq:wechsler} is a poor approximation to the MAH of a
significant number of systems.

\subsubsection{NFW prescription}
\label{sssec:nfw}

Finally, we consider NFW's proposal to identify the formation redshift
with the epoch when $50\%$ of the halo is contained in progenitors
more massive than certain fraction, $f$, of the final halo mass. NFW
propose $f=0.01$ in their original work in order to match the
mass-concentration relation using the extended Press-Schechter
formalism, but this fraction is too low to allow for an accurate
estimate in N-body simulations. As a compromise, we adopt $f=0.1$ for
the results shown here.

\subsubsection{Comparison between prescriptions}
\label{sssec:comp}

Note that the three prescriptions described above are based on
different features of the halo merger trees. While NFW looks at the
mass spectrum of clumps containing half of the final halo mass,
the other methods consider just the MAH of the most massive
progenitor.  The Z03 prescription depends on the slope of
the scaling relation $\log_{10}M_{\rm s}$ vs. $\log_{10}r_{\rm s}$ in
the slow accretion phase while the W02 recipe fits the
whole MAH with a single slope. 

\begin{figure}
\includegraphics[width=8cm]{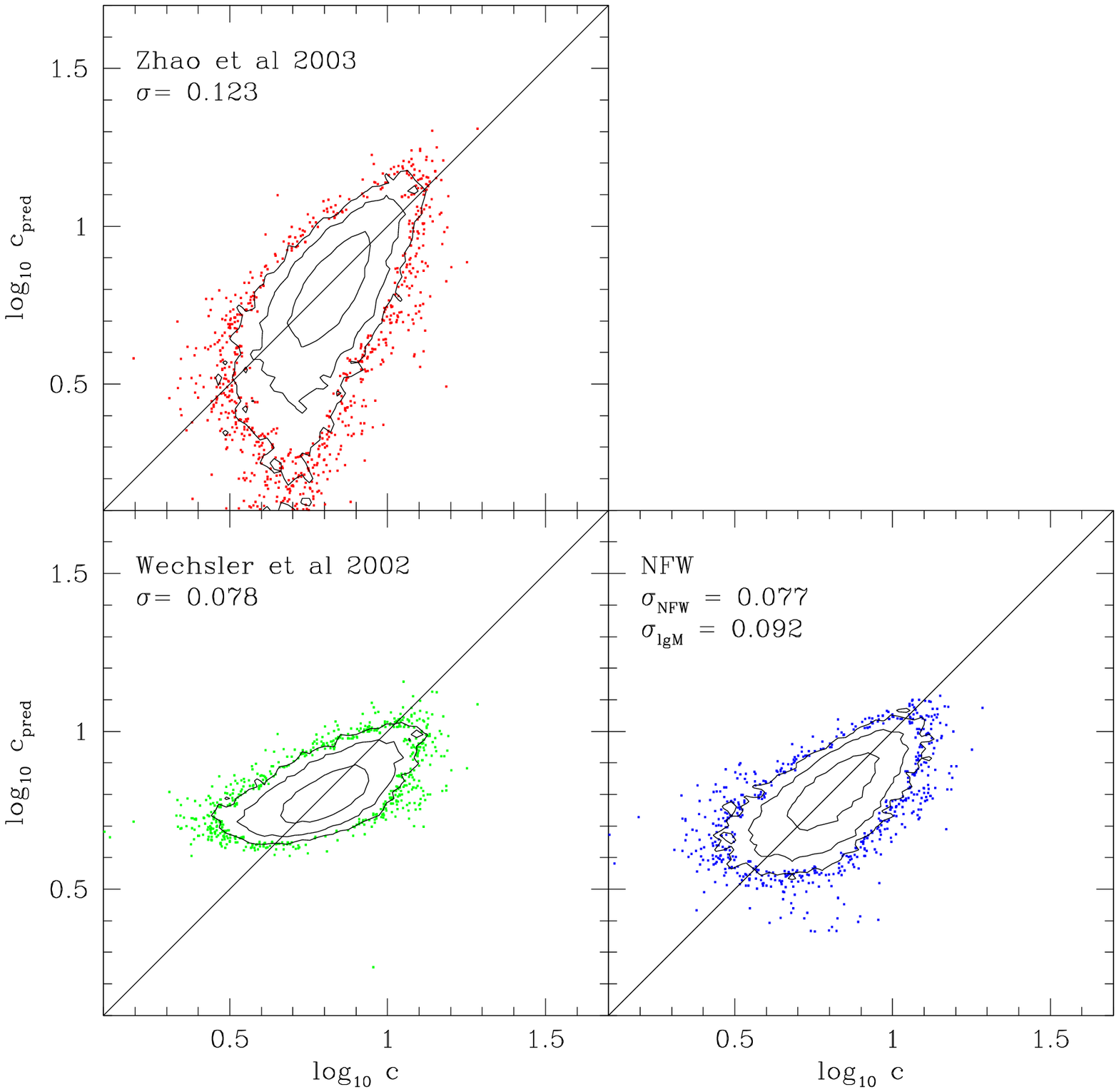}
\caption{A comparison of the measured concentrations with those
predicted by the \protect\cite{zhao03a} (top), \protect\cite{wechsler02}
(bottom left) and NFW (bottom right) prescriptions.  The plotted
contours enclose 65\%, 95\% and 99\% of the haloes with the remaining
1\% plotted as points. The $\sigma$-values indicated in each panel
give the rms scatter in the prediction $\langle (\log_{10} (c/c_{\rm
pred}))^2 \rangle^{1/2}$, while $\sigma_{\rm lgM}$ is the
corresponding rms scatter about the mass-concentration relation for
the same set of haloes.}
\label{fig:predictions}
\end{figure}

In spite of these differences, Fig.~\ref{fig:predictions} shows that
all three procedures yield concentrations that correlate reasonably
well with the measured values. The rms scatter between prediction and
measurement is indicated in each panel. It is smallest (marginally)
for the NFW prescription, but even in this case it only reduces the scatter
in the mass-concentration relation (Fig.~\ref{fig:massc}) from
$\sigma_{\rm lgM}=0.092$ to $\sim 0.077$.  Thus it
only accounts for about 30\% ($\vert (\sigma_{\rm lgM}^2-\sigma_{\rm
NFW}^2)/\sigma_{\rm lgM}^2 \vert$) of the variance in the
mass-concentration relation. 

The W02 prescription does similarly well by this
measure, but the slope of the $c_{\rm pred}$--$c_{\rm measured}$
relation is a bit too shallow.  The Z03 prediction has more
scatter, but this is entirely due to a tail of haloes for which it
predicts very low concentrations.  We conclude that all three methods
predict concentrations that correlate well with the measured values,
but none of them is able to fully account for the scatter in the
mass-concentration relation.

%%%%%%%%%%%%%%%%%%%%%%%%%
%%%%%%%%%%%%%%%%%%%%%%%%%

\section{Summary}
\label{sec:conc}

We use the {\it Millennium Simulation} to examine the structural
parameters of dark matter haloes formed in the $\Lambda$CDM
cosmogony. The large volume probed by the {\sl MS}, together with its
unprecedented numerical resolution, allow us to probe confidently the
mass profiles of haloes spanning more than three decades in mass. Our
main conclusions may be summarised as follows.

\begin{itemize}

\item As in earlier studies, we find that the mass profile of
dynamically {\it relaxed} haloes are well approximated by the
two-parameter NFW profile. We find that at least $1000$ particles are
needed in order to obtain unbiased estimates of the distribution of
halo concentrations and illustrate a number of potential pitfalls that
arise from analysing poorly-resolved haloes, or from including in the
sample haloes manifestly out of equilibrium.

\item We study the correlation between the NFW fit parameters, which
we express in terms of the halo mass and a concentration
parameter. These results extend previous studies to much larger halo
masses than hitherto reported in the literature. Combining our results
with those of \cite{maccio07}, we find that a single power law
reproduces the mass-concentration relation for over six decades in
mass. These results are in reasonable, albeit not perfect, agreement
with the predictions of the \cite{eke01} and NFW models. The model of \cite{bullock01} fails at large masses, and predicts concentrations at
least a factor of $\sim 2$ too low for $M\sim 10^{15} \, h^{-1}
M_{\odot}$ haloes.

\item The dependence of concentration on mass, while well established,
is weak, and of equal importance is the broad scatter in concentration
at fixed mass. The distribution of concentrations at given mass is
well fitted by a lognormal function where {\it both} the mean and the
dispersion decrease with increasing halo mass. These results allow us
to estimate in detail the abundance of haloes with unusually low or
unusually high concentration, providing a well-defined prediction that
may be used to interpret observations of objects of unusual density,
such as highly-effective cluster lenses or galaxies with haloes of
anomalously low density.

\item We find that, once {\it unrelaxed} haloes are excluded, there is
no significant correlation between halo spin and concentration,
contrary to the results of \cite{bailin05}.

\item We have searched for several ways to account for the large
dispersion in concentrations at given mass. The scatter in
concentrations seems to arise largely due to variations in the
formation time. We examined various plausible definitions of the
formation time, and find that concentrations are best predicted by
formation times defined taking into account the collapse history of
the full spectrum of progenitors rather than the evolution of the
single most massive progenitor.

\item We compare the schemes of \cite{zhao03a}, \cite{wechsler02}, and a
variant of the NFW prescription, and find that, while all three show a
strong correlation between the predicted and measured concentrations,
considerable scatter remains. In fact, none of these models is able to
account for more than 30\% of the intrinsic variance in the
mass-concentration relation. It appears as if a large fraction of the
scatter is truly stochastic or, else, dependent on aspects of the halo
merger history that are not probed by these simple schemes.

\end{itemize}

\section*{acknowledgements}
The simulation used in this paper was carried out as part of the
programme of the Virgo Consortium on the Regatta supercomputer of the
Computing Centre of the Max-Planck Society in Garching. AFN would like
to thank for the hospitality of ICC during the year 2005 and the LENAC
network for financial support. AFN would also like to acknowledge John
Helly for his help with the merger trees. JFN acknowledges support
from Canada's NSERC, as well as from the Leverhulme Trust and the
Alexander von Humboldt Foundation. PB acknowledges the receipt of a
PhD studentship from the UK Particle Physics and Astronomy Research
Council.

\begin{table*}
\caption{Parameters of the log-normal fits (eq.~\ref{eq:lognormal}) to
the distribution of concentrations as a function of mass for bins
shown in Fig.~\ref{fig:massc}.  The first two columns are the halo
mass, $M_{200}$, and the number of haloes in each mass bin. The
following two columns denote the median and dispersion in the
logarithm of the concentration; $\langle \log_{10} c_{200} \rangle$,
and $\sigma_{\log_{10} c}$ for the {\it relaxed} halo sample. The last
three columns are the fraction of unrelaxed haloes, $f_{\rm unrel}$, as
well as the median and dispersion in the logarithm of the
concentration for that sample.}
\begin{tabular}{ccccccc}
\hline
$\log_{10} M_{200}$ & 
${\rm N}_{\rm haloes}$ & 
$\langle \log_{10} c \rangle$ & 
$\sigma_{\log_{10}c}$ & 
$f_{\rm unrel}$ & 
$\langle \log_{10} c \rangle$ & 
$\sigma_{\log_{10} c}$ \\
$[h^{-1}\, M_\odot]$ & &[relaxed] & [relaxed]& &[unrelaxed] & [unrelaxed]\\
\hline

11.875 - 12.125 & 911 & 0.920 & 0.106 & 0.205 & 0.683 & 0.147 \\
12.125 - 12.375 & 721 & 0.903 & 0.108 & 0.171 & 0.658 & 0.150 \\
12.375 - 12.625 & 403 & 0.881 & 0.099 & 0.199 & 0.646 & 0.139 \\
12.625 - 12.875 & 165 & 0.838 & 0.101 & 0.229 & 0.605 & 0.158 \\
12.875 - 13.125 & 13589 & 0.810 & 0.100 & 0.263 & 0.603 & 0.136 \\
13.125 - 13.375 & 9194 & 0.793 & 0.099 & 0.253 & 0.586 & 0.140 \\
13.375 - 13.625 & 4922 & 0.763 & 0.095 & 0.275 & 0.566 & 0.142 \\
13.625 - 13.875 & 2474 & 0.744 & 0.094 & 0.318 & 0.543 & 0.140 \\
13.875 - 14.125 & 1118 & 0.716 & 0.088 & 0.361 & 0.531 & 0.131 \\
14.125 - 14.375 & 495 & 0.689 & 0.095 & 0.383 & 0.510 & 0.121 \\
14.375 - 14.625 & 191 & 0.670 & 0.094 & 0.370 & 0.490 & 0.133 \\
14.625 - 14.875 & 49 & 0.635 & 0.091 & 0.484 & 0.519 & 0.121 \\
14.875 - 15.125 & 8 & 0.664 & 0.061 & 0.578 & 0.493 & 0.094 \\

\hline
\end{tabular}
\label{tab:fits}
\end{table*}

\setlength{\bibhang}{2.0em}

\end{document}